\documentclass[12pt]{article}
\usepackage{makeidx}
\usepackage{float}
\usepackage{booktabs,threeparttable,siunitx}
\usepackage{times}
\usepackage{microtype}
\usepackage{amssymb}
\usepackage{amsmath}
\numberwithin{equation}{section}
\usepackage{setspace}
\usepackage[round,authoryear,comma]{natbib}
\usepackage{hyperref}
\usepackage{graphicx}
\usepackage{xcolor}

\usepackage{multirow}
\usepackage{array}
\usepackage{algorithm}
\usepackage{algpseudocode}
\usepackage{placeins}
\hypersetup{
	colorlinks   = true,
	urlcolor     = blue,
	linkcolor    = blue,
	citecolor    = blue
}

\newtheorem{theorem}{Theorem}[section]
\newtheorem{corollary}{Corollary}[section]

\newtheorem{assumption}{Assumption}
\newtheorem{lemma}{Lemma}[section]
\newtheorem{proposition}{Proposition}[section]
\newtheorem{remark}{Remark}[section]
\newenvironment{proof}[1][Proof]{\noindent \textbf{#1.} }{\hfill\rule{0.5em}{0.5em}}
\parskip=\medskipamount
\newcolumntype{C}[1]{>{\centering\arraybackslash}m{#1}}
\newcommand{\XD}{2.483}

\newcommand{\XcirK}{0.144}
\newcommand{\XcirM}{0.0089}
\newcommand{\XcirS}{0.054}
\newcommand{\XcirFL}{0.00258}
\newcommand{\XcirFR}{0.00294}
\newcommand{\XcirN}{296}
\newcommand{\Xrzero}{0.02}
\newcommand{\XnRuns}{100}

\newcommand{\XnCorner}{3}

\newcommand{\XnNonCorner}{97}
\newcommand{\XlpLS}{24.78}
\newcommand{\XkaLS}{-0.697}
\newcommand{\XmultLS}{8.70}
\newcommand{\XlamHatTls}{2.19}
\newcommand{\XlpSpreadLS}{0.059}

\newcommand{\XrmseMin}{1.418}
\newcommand{\Xalpha}{83.2}

\newcommand{\XsseTarget}{40.71}

\newcommand{\XlpStar}{2.549}
\newcommand{\XkaStar}{1.037}
\newcommand{\XetaStar}{1.267}
\newcommand{\XmultStar}{9.20}
\newcommand{\XKStar}{0.3275}
\newcommand{\XrmseStar}{1.504}
\newcommand{\XitStar}{27}
\newcommand{\XdevRef}{$1.7\times10^{-5}$}

\newcommand{\XqPrice}{99.694}
\newcommand{\XpPrice}{113.666}
\newcommand{\XactPrice}{107.876}
\newcommand{\XpremZero}{7.876}
\newcommand{\XrelLp}{3.01}
\newcommand{\XrelKa}{0.980}
\newcommand{\XresidDec}{6.25}
\newcommand{\XresidMaxOther}{0.80}
\newcommand{\XnFeas}{17}

\newcommand{\XmultMinEx}{6.58}
\newcommand{\XmultMaxEx}{10.45}

\newcommand{\XmultZero}{8.87}
\newcommand{\XEL}{0.245}
\newcommand{\XspreadEL}{8.81}
\newcommand{\XmultDec}{18.26}
\newcommand{\XlpMinDW}{1.741}
\newcommand{\XlpMaxDW}{5.366}
\newcommand{\XkaMinDW}{0.913}
\newcommand{\XkaMaxDW}{1.224}
\newcommand{\XfsMin}{53.5}
\newcommand{\XfsMax}{63.7}
\newcommand{\Xdevone}{$9.7\times10^{-7}$}
\newcommand{\XdevOcc}{0.084}
\newcommand{\XitDWmin}{76}
\newcommand{\XitDWmax}{97}
\newcommand{\XpiMin}{0.357}
\newcommand{\XpiMax}{12.620}

\newcommand{\XmcN}{1}
\newcommand{\XmcZ}{0.98}
\newcommand{\XoccAgg}{0.0459}
\newcommand{\XoccAggP}{0.0051}
\newcommand{\XlamHatT}{0.302}
\newcommand{\XlamHatTbound}{12.6}
\newcommand{\XuniqN}{480}

\newcommand{\XliMult}{4.30}
\newcommand{\XliFeas}{14}

\newcommand{\XcirMult}{9.15}

\newcommand{\XdAltMult}{5.13}
\newcommand{\XzLoMult}{6.13}
\newcommand{\XzHiMult}{18.45}
\newcommand{\XvMin}{7.66}
\newcommand{\XvMax}{11.57}
\newcommand{\XfsRobMin}{49.2}
\newcommand{\XfsRobMax}{71.6}
\newcommand{\XalphaMultMin}{4.66}
\newcommand{\XalphaMultMax}{10.03}

\newcommand{\XexRmse}{0.335}

\graphicspath{{figures/}}

\begin{document}

\title{An Application of Geometric Riemannian Pricing Algorithm for Valuing Catastrophe Bonds in Incomplete Markets: Market-Consistent Financial Entropy and the Split of Frequency and Severity Risk Premia}
\author{Saeid Safarveisi\footnote{Corresponding author} \\ \small Actuarial Research Group, AFI, Faculty of Economics and Business,
KU Leuven, Leuven, Belgium \\ \small \texttt{saeid.safarveisi@kuleuven.be}}
\date{Version: \today}
\maketitle
\hrule
\begin{abstract}
\noindent This paper presents a framework for pricing catastrophe (CAT) bonds in incomplete markets based on Riemannian geometric optimization of financial entropy. CAT bonds, which transfer extreme event risk from insurers to capital markets, are notoriously difficult to price due to market incompleteness and limited observed data. We formulate the selection of the pricing measure as a maximum financial entropy problem over a finite-dimensional manifold of equivalent measures, in which the non-hedgeable market prices of catastrophe frequency and severity risk are the coordinates, while discounting is fixed by the default-free term structure. We show that, without market information, maximum financial entropy simply returns the physical measure, and we therefore combine the entropy criterion with observed secondary-market spreads, both pooled over the life of the bond and date by date. For the compound Poisson--generalized Pareto loss model with a Wang distortion, the financial entropy and the associated Riemannian metric are available in closed form. We also show that CAT bond prices depend on the frequency and severity premia essentially only through the risk-neutral trigger intensity, so market data alone cannot separate the two; the entropy criterion resolves precisely this direction. The framework is illustrated with a hypothetical earthquake CAT bond calibrated to quarterly market-indicated spreads of Golden State Re II Ltd.\ (Series 2014-1). Market prices imply a risk-neutral trigger intensity of $\XmultMinEx$--$\XmultMaxEx$ times its physical value ($\XmultStar$ for the pooled measure, and $\XmultDec$ at the end of 2015). This risk multiple is the model-consistent counterpart of the market's spread-to-expected-loss ratio ($\XmultZero$ against $\XspreadEL$ at issuance). The minimal-entropy measure attributes $\XfsMin$--$\XfsMax\%$ of the log risk multiple to frequency; this split is a property of the selection principle and the trigger level rather than of the market data. We also show that setting the coupon equal to the quoted spread, a common simplification, understates the risk multiple by about one half.

\end{abstract}
\hrule
\newpage
\section{Introduction}

\noindent In the traditional insurance framework, insurers transfer portions of their risk portfolios to (re)insurers in order to improve solvency and reduce exposure to large losses. Traditional reinsurance capacity is, however, limited relative to the losses that major natural catastrophes can generate. Insurance-linked securities (ILS), and catastrophe (CAT) bonds in particular, transfer such risks to capital markets. Since their introduction in the 1990s CAT bond issuance has grown substantially, although not uniformly.

\vspace{0.2cm}
\noindent The evolution of the CAT bond market has been accompanied by extensive research aimed at developing valuation methodologies suited to incomplete markets. Tang and Yuan \citeyearpar{Ref39} provide a comprehensive survey of CAT bond pricing approaches, categorising existing models into six main frameworks: zero risk premium, arbitrage pricing theory (APT), probability transform, econometric methods, indifference pricing, and two-step valuation. Each framework offers a different perspective on the treatment of catastrophe risk and the construction of risk-adjusted pricing measures.

\vspace{0.2cm}
\noindent One such framework, as mentioned above, is the zero risk premium approach. In an arbitrage-free but incomplete market setting, this approach treats catastrophe risks as zero-beta assets with zero risk premium, following Merton's \citeyearpar{Ref34} insight. Cox and Pedersen \citeyearpar{Ref11} applied this principle to construct a pricing measure for catastrophe events, and several authors have subsequently extended their model. Shao et al.\ \citeyearpar{Ref37} generalise Cox and Pedersen's construction by jointly modelling financial and actuarial risks. Lee and Yu \citeyearpar{Ref31} propose a contingent-claim framework incorporating moral hazard, basis risk, and default risk. Baryshnikov et al.\ \citeyearpar{Ref3} and Burnecki and Kukla \citeyearpar{Ref8} study CAT bond pricing within compound doubly Poisson process frameworks, while Vaugirard (\citeyear{Ref42}a,b) introduces an additional diffusion component to capture randomness in both catastrophe processes and interest rates.

\vspace{0.2cm}
Another influential stream of research is based on distorted probability measures: Venter \citeyearpar{Ref41} argues that catastrophe risks should be valued under a risk-adjusted probability distribution formed by distorting the physical measure. This distortion-based viewpoint has been widely adopted in the CAT bond literature; see Tang and Yuan \citeyearpar{Ref39}, Blanchet et al.\ \citeyearpar{Ref5}, and Godin et al.\ \citeyearpar{Ref26}, among many others. A further class of valuation methods arises from utility-based pricing, where subjective risk aversion is captured through a utility function, as introduced by Hodges and Neuberger \citeyearpar{Ref28}. Related developments include the work of Barrieu and Louberg\'{e} \citeyearpar{Ref2}, Karagiannis et al.\ \citeyearpar{Ref44}, Egami and Young \citeyearpar{Ref19}, and Ma \citeyearpar{Ref99996}.

\noindent Complementary modelling approaches have also been proposed. Ma and Ma \citeyearpar{Ref33} develop a mixture model to approximate the distribution of aggregate losses, and Shao et al.\ \citeyearpar{Ref38} extend this framework by incorporating dependence between claim sizes and inter-arrival times using a semi-Markov structure. More recently, Safarveisi et al.\ \citeyearpar{Ref8889} investigated the impact of the inter-arrival time distribution on CAT bond pricing by considering a renewal compound process that preserves its distributional structure under the measure change.

\vspace{0.2cm}
\noindent While many studies have examined risk modelling and measure transformation in CAT bond pricing, two questions are less studied. First, \emph{which} element of the set of market-consistent measures should be used when observed prices do not pin it down. Second, \emph{whether} the risk premia being calibrated are identified by the available prices at all. The inherent incompleteness of the CAT bond market---where no unique pricing measure exists---raises the question of how to select an appropriate measure for pricing purposes. A natural answer is the minimal entropy martingale measure, where an equivalent measure $\mathbb{Q}$ is selected to be as close as possible to the physical measure in the sense of relative entropy, subject to the prices of traded securities; see Dhaene et al.\ \citeyearpar{Ref1901}. Closely related is the calibration of pricing models by relative-entropy minimisation subject to observed prices, see Avellaneda et al.\ \citeyearpar{RefAv97} and Cont and Tankov \citeyearpar{RefCT04}. Within a Bayesian framework, Domfeh et al.\ \citeyearpar{RefDomfeh22} combine a hierarchical Bayesian collective risk model with a CIR interest-rate model and convert the simulated physical scenarios into a single risk-neutral distribution by maximum entropy, subject to an issue-price constraint. We instead work with a parametric continuous-time family of pricing measures, use the financial entropy of Gao et al.\ \citeyearpar{Ref24001}, calibrate to a time series of secondary-market quotes, and characterise which components of the risk premium these prices identify. We adopt the more general notion of financial entropy of Gao et al.\ \citeyearpar{Ref24001}, which covers utility functions beyond the exponential, and use their Riemannian geometric construction to solve the resulting optimisation problem in a continuous-time setting. In this paper we work with the exponential-utility case, in which financial entropy reduces to relative entropy.

\vspace{0.2cm}
\noindent Two observations shape our formulation. First, if the entropy criterion is applied without any market information, its solution is the physical measure itself, so that the risk premium vanishes (Proposition~\ref{prop:degenerate}). Market prices must therefore enter the problem, and the hedgeable interest-rate premium should be fixed by the default-free term structure rather than by entropy, as in Gao et al.\ \citeyearpar{Ref24001}. Second, for a CAT bond that pays on the first crossing of a loss threshold, prices depend on the market prices of frequency risk $\lambda^\prime$ and severity risk $\kappa$ essentially only through the risk-neutral trigger intensity (Proposition~\ref{prop:ident}). Direct (na\"ive) calibration of both premia is therefore close to unidentified, whatever the number of observed spreads, and the entropy criterion is exactly what selects the split between the two.

\vspace{0.2cm}
\noindent Our empirical application concerns a hypothetical earthquake CAT bond calibrated to the Golden State Re II Ltd.\ (Series 2014-1) notes, using 18 quarter-end secondary-market spreads reported by Lane Financial L.L.C.\ from September 2014 to December 2018, i.e.\ over the life of the bond. We use this market information in two complementary ways. In the \emph{pooled} analysis, a single market-price-of-risk vector $\theta=(\lambda^\prime,\kappa)$, comprising the non-hedgeable frequency ($\lambda^\prime$) and severity ($\kappa$) components, is selected for the whole observation period by combining the entropy criterion with all quarterly spreads. In the \emph{date-wise} analysis, the entropy is maximised separately at each quarter, subject to reproducing that quarter's price exactly. The pooled analysis asks which single pricing measure is most consistent with the market over the life of the bond; the date-wise analysis asks how the market's pricing of catastrophe risk evolved from quarter to quarter. Throughout, discounting uses the observed Treasury curve, so that the interest-rate premium never enters the entropy problem.

\vspace{0.2cm}
\noindent In implementing the pooled Riemannian optimisation, a central numerical question is the selection of a reliable initial value for $\theta$. Direct (na\"ive) calibration of $\theta$ by minimising pricing errors (squared deviations between model CAT bond values and par at the observed spreads) is hampered by two structural features of the available data: (i) \emph{sparsity}---the number of observed quarterly spread points for a given issue is small; and (ii) \emph{entanglement}---the two catastrophe premia enter prices through (almost) the same combination, so that the sum-of-squares surface is nearly flat along a whole curve of parameter values. Its minimum is then determined by second-order effects of the model and, in our application, lies far from the physical measure. We therefore use the na\"ive least-squares estimate only as a starting value: the entropy term of the pooled criterion selects the point on the nearly flat curve that is closest to $\mathbb P$, and because its maximiser is unique (verified on a grid), any reasonable initialisation leads to the same pricing measure and can only affect the computing time.

\begin{figure}[htbp]
\centering
\includegraphics[width=\linewidth]{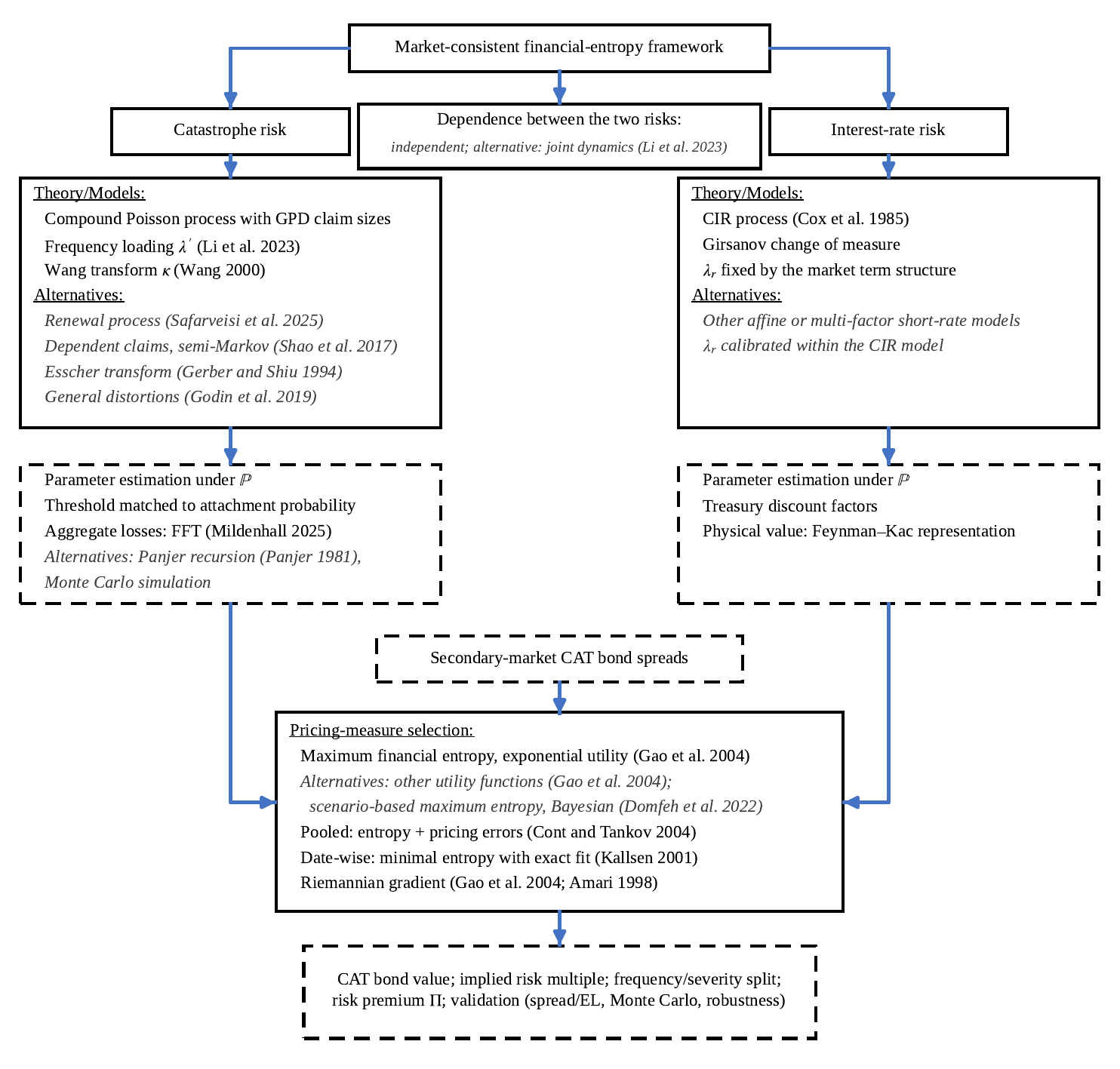}
\caption{Overview of the valuation framework. Solid boxes contain the models and methods; dashed boxes contain estimation, data and outputs. Alternatives that can be used within the same framework are shown in italics.}
\label{fig:framework}
\end{figure}

\vspace{0.2cm}
\noindent Building on the geometric algorithm of Gao et al.\ \citeyearpar{Ref24001}, the measure change of Li et al.\ \citeyearpar{Ref989800} and the entropy-regularised calibration of Cont and Tankov \citeyearpar{RefCT04}, we contribute as follows. (i) We show that maximum financial entropy is uninformative for CAT bonds without market constraints, and we formulate market-consistent versions, pooled and date-wise, with discounting fixed by the term structure. (ii) We prove that CAT bond prices identify, essentially, only the risk-neutral trigger intensity, which explains the difficulty of direct calibration; the entropy criterion selects the split of this intensity into frequency and severity loadings, and the selected measures lie on a single minimal-entropy path determined by the identified intensity. (iii) We derive closed-form expressions for the financial entropy and the Riemannian metric of the measure family, which make the geometric algorithm transparent and cheap. (iv) We provide an empirical illustration, including the time series of implied trigger intensities and premium splits. We show that the implied risk multiple is the model-consistent counterpart of the market's spread-to-expected-loss ratio, and that the convention used to translate quoted spreads into prices matters: setting the coupon equal to the spread understates the multiple by about one half. Figure~\ref{fig:framework} summarises the framework and the choices made in this paper.

\vspace{0.2cm}
The remainder of the paper is organized as follows. Section 2 presents the model setup and the assumptions for insurance and financial risks, assuming independence between the two, and derives the Radon--Nikodym derivative used for pricing purposes. Section 3 introduces the valuation framework, the pooled and date-wise selection of the pricing measure based on financial entropy and market information, and the associated algorithms for a hypothetical CAT bond, together with the identification results. Section 4 describes how to compute the CAT bond value under the physical measure for comparison. Section 5 provides the numerical analysis, including the pooled and date-wise analyses and robustness checks. Finally, Section 6 concludes the paper.

\section{Model setup and assumptions}
Let $(\Omega, \mathcal{F}, \{\mathcal{F}_{t}\}_{0\leq t\leq T}, \mathbb{P})$ be a filtered probability space accommodating all sources of uncertainty arising from different risks, where $\Omega$ is the sample space, including all the possible paths taken by the respective variables; $\{\mathcal{F}_{t}\}_{0\leq t\leq T}$ is an increasing filtration representing the information available up to and including time $t$, with $T$ as the time horizon; $\mathcal{F}=\sigma(\cup_{0\leq t\leq T}\mathcal{F}_{t})$; and $\mathbb{P}$ denotes the physical probability measure defined on $\mathcal{F}$. In the sequel, we describe the models driving the interest rate and the loss index, which are critical to our CAT bond valuation framework.

\subsection{Insurance risk}
Let the accumulated losses over a specified period be modelled by a compound Poisson process involving two key components: frequency and severity. The frequency component represents the number of earthquake events striking the insured region (as stipulated by the bond contract) and is governed by a Poisson process $\{N(t):t\geqslant 0\}$ with intensity parameter $\lambda$. The severity component, representing the (modelled) loss caused by each earthquake, is captured by a sequence of independent and identically distributed positive random variables $\{X_k : k\in \mathbb{N}\}$ with the common continuous distribution function $F(x) = \mathbb{P}(X_k \leqslant x)$ and survival function $\bar F=1-F$. Consequently, the aggregate loss process $\{L_t: t\in [0,T]\}$ is defined as $L_t = \sum_{k=1}^{N(t)}X_k$, where $\{X_k\}$ and $N$ are assumed to be independent.

\subsection{Financial risk}
Let the risk-free interest rate be modelled by the Cox--Ingersoll--Ross (CIR) model (Cox et al.\ \citeyear{Ref10}a), also known as the mean-reverting square-root process:
\begin{eqnarray}\label{ref3}
dr(t) = \bar{\theta}_r (\bar{m}_r - r(t))dt + \sigma_r \sqrt{r(t)}dW(t),
\end{eqnarray}
under measure $\mathbb{P}$, where $W(t)$ is a Brownian motion, $\bar{\theta}_r>0$ is the speed of mean reversion, $\bar{m}_r>0$ is the long-run interest rate mean, and $\sigma_r>0$ is the volatility parameter. The Feller condition (Feller \citeyear{Ref21}), $2\bar{m}_r \bar{\theta}_r\geq \sigma_r^2$, makes $r(t)$ almost surely strictly positive. We refer the reader to Burnecki et al.\ \citeyearpar{Ref7} and Fergusson \citeyearpar{Ref22} for different extensions of the CIR model.

\subsection{Traded assets and market data}
\begin{assumption}[Traded assets]\label{ass:traded}
The following are traded: (a) the money-market account $B(t)=\exp\{\int_0^t r(s)\,ds\}$; (b) default-free zero-coupon bonds $P(t,T^\prime)$ for the relevant maturities, whose prices $P^{\mathrm{mkt}}(t,T^\prime)$ are observed (Treasury curve); and (c) the CAT bond, whose secondary-market prices $V^{\mathrm{mkt}}_{t_k}$ are observed at dates $t_0<t_1<\dots<t_K$. The market is free of arbitrage, so that the set $\mathcal M$ of equivalent martingale measures for the discounted traded prices is non-empty.
\end{assumption}
\begin{remark}[Secondary market]\label{rem:secondary}
The prices targeted here are \emph{secondary-market} prices of a bond already in issue. The coupon spread of a CAT bond is fixed at issuance, but the bond subsequently trades at prices that reflect current market conditions, such as the supply of and demand for catastrophe risk, and therefore need not coincide with its issue terms; the outstanding CAT bond market as a whole trades at a premium or a discount to its issue value depending on market conditions (Lane and Beckwith \citeyear{RefLane}). Market information on traded bonds is typically reported as a \emph{market-indicated spread}, i.e.\ the spread over the risk-free rate at which the bond currently trades, rather than as a price. For Golden State Re II, for instance, the spread at issue is 2.20\%, while the quarter-end indications vary between 1.65\% and 4.56\% over 2014--2018. Since our pricing model produces prices, each quoted spread must be translated into a market price before calibration; Section~\ref{ref54} explains the convention we use and shows that this choice matters.
\end{remark}

\subsection{Derivation of pricing measure}\label{section2.3}
Following Li et al.\ \citeyearpar{Ref989800} and in accordance with the model framework described above, we introduce the pricing measure $\mathbb{Q}$ via the following construction. Define
\begin{eqnarray}
Z_{1,t} &:=& \exp\left\{\int_{0}^{t}\gamma(s)dW(s) - \frac{1}{2}\int_{0}^{t}\gamma^2(s)ds\right \},\\
Z_{2, t} &:=&  e^{(\lambda - \hat{\lambda})t }\prod_{i=1}^{N(t)}\left( \frac{\hat{\lambda}}{\lambda }g(X_i)\right), \qquad 0\leq t \leq T,
\end{eqnarray}
and let
\begin{eqnarray}
\frac{d\mathbb{Q}}{d\mathbb{P}} := Z_T = Z_{1,T}\,Z_{2, T}  \quad \text{on} \;\mathcal{F}_T,
\end{eqnarray}
where $Z_T$ is the Radon--Nikodym derivative, $Z_t=\mathbb E^{\mathbb P}[Z_T|\mathcal F_t]$ is a $\mathbb{P}$-martingale, and $L_t$ is a compound Poisson process with frequency intensity $\hat{\lambda}$ and severity density $\hat{f}$ under $\mathbb{Q}$. The function $g(x) = \frac{d\hat{F}}{dF}(x)$, for $x$ in the support of $X$, is positive, measurable and satisfies $\mathbb E^{\mathbb P}[g(X)]=1$. Throughout, we assume that $W(t)$ and $L_t$ are independent under $\mathbb{P}$. Because the density factorises into a factor depending only on $W$ and a factor depending only on $(N,\{X_k\})$, the independence also holds under $\mathbb{Q}$. Furthermore, $Z_{1,t}$ corresponds to the usual change-of-measure kernel in Girsanov's theorem, which implies that
\begin{eqnarray}
W^{*}(t) = W(t) - \int_{0}^{t}\gamma(s)ds \quad \text{or equivalently} \quad dW^{*}(t) = dW(t) -\gamma(t)dt,
\end{eqnarray}
is a Brownian motion under $\mathbb{Q}$. As a result, under the pricing measure $\mathbb{Q}$, the short rate $r(t)$ follows a modified CIR process
\begin{eqnarray}\label{eq:cirQ}
dr(t) = \theta_{r}^{*}(m^{*}_{r} - r(t))dt + \sigma_{r}\sqrt{r(t)}dW^{*}(t),
\end{eqnarray}
where, with the specification of $\gamma$ in \eqref{eq:gamma} below,
\begin{eqnarray}\label{eq:thetastar}
\theta_{r}^{*} = \bar{\theta}_r - \frac{\gamma(t)}{\sqrt{r(t)}}\sigma_r=\bar\theta_r+\lambda_r , \qquad m^{*}_{r} = \frac{\bar{\theta}_r\bar{m}_r}{\theta_{r}^{*}}.
\end{eqnarray}
We require $\theta_r^*>0$, so that the short rate remains mean-reverting under $\mathbb Q$. Since $\theta_r^*m_r^*=\bar\theta_r\bar m_r$, the Feller condition holds under $\mathbb Q$ whenever it holds under $\mathbb P$.

\noindent Next, we define the market price of frequency risk, $\lambda^\prime>-1$, as
\begin{eqnarray}
    \lambda^\prime = \frac{\hat{\lambda}}{\lambda} -1, \qquad\text{and write}\qquad \eta_f:=\ln(1+\lambda^\prime),
\end{eqnarray}
where $\eta_f$ is the log-scale frequency parameter. We embed the market price of severity risk, $\kappa$, in the Wang distortion transform\footnote{We use the Wang transform (Wang \citeyear{RefWang00}) for tractability: the density ratio $g$ is explicit, and the relative entropy and the Riemannian metric of the resulting family of pricing measures are available in closed form. Other choices are possible within the same framework. A commonly used alternative is the Esscher transform (Gerber and Shiu \citeyear{RefGS94}), $g(X)=e^{\kappa X}/\mathbb{E}^{\mathbb{P}}[e^{\kappa X}]$, which requires a finite moment generating function of the claim size and can be used with lighter-tailed severity distributions. More general classes of distortion operators, such as those of Godin et al.\ \citeyearpar{Ref26}, may be better suited to the heavy-tailed losses typical of catastrophe events. The choice of the Wang transform is not a limitation of the method: when closed forms are not available, the financial entropy, the Riemannian metric and the prices can be computed by simulation, and the pricing algorithm applies unchanged.}
\begin{eqnarray}\label{eq:wang}
\hat F(x)=\Phi\big(\Phi^{-1}(F(x))-\kappa\big),\nonumber\\
g(X) = \frac{\phi(\Phi^{-1}(F(X)) -\kappa)}{\phi(\Phi^{-1}(F(X)))}=\exp\Big\{\kappa\,\Phi^{-1}(F(X))-\frac{\kappa^2}{2}\Big\},
\end{eqnarray}
where $\Phi$ and $\phi$ are the standard normal distribution function and density. With this (standard) sign convention, $\kappa>0$ shifts probability mass towards larger losses, i.e.\ it is a positive severity loading. Equivalently, the normal scores $Z_i:=\Phi^{-1}(F(X_i))$ are $N(0,1)$ under $\mathbb P$ and $N(\kappa,1)$ under $\mathbb Q$. Finally, the market price of the interest rate (diffusion) risk is specified through
\begin{eqnarray}\label{eq:gamma}
    \gamma(t)=\frac{-\lambda_r}{\sigma_r}\sqrt{r(t)} .
\end{eqnarray}
Combining these results, the Radon--Nikodym derivative can be expressed explicitly as
\begin{eqnarray}\label{ref51001}
\displaystyle \frac{d\mathbb{Q}}{d\mathbb{P}} =  \exp\left\{-\int_{0}^{T} \frac{\lambda_r}{\sigma_r}\sqrt{r(s)}dW(s) - \frac{1}{2}\int_{0}^{T}\frac{\lambda_r^2}{\sigma_r^2}r(s)ds -\lambda \lambda^\prime T\right \} (1+\lambda^\prime)^{N(T)}\prod_{i=1}^{N(T)}g(X_i).
\end{eqnarray}
We assume that $Z_{1}$ is a true $\mathbb P$-martingale; for this specification the $\mathbb Q$-dynamics \eqref{eq:cirQ} are again of CIR type, and the assumption is standard. $Z_2$ is a true martingale because $\mathbb E^{\mathbb P}[g(X)]=1$ and $N$ is Poisson.

\begin{remark}[Hedgeable premia]\label{rem:hedge}
Default-free zero-coupon bonds are traded, so the interest-rate premium $\lambda_r$ is \emph{hedgeable}. In the main analysis we therefore discount with the observed Treasury discount factors $P^{\mathrm{mkt}}(t_k,t_j)$. Because $W$ and $L$ are independent under every $\mathbb Q$ in our family, this is equivalent to fixing $\lambda_r$ (or any other interest-rate model) consistently with the curve, and $\lambda_r$ never has to be estimated. The CIR version with $\lambda_r$ calibrated to the curve through relation~\eqref{ref101} (Appendix A) is used only as a robustness check. In either case $\lambda_r$ is \emph{not} selected by entropy. The two jump components $\lambda^\prime$ and $\kappa$ are \emph{non-hedgeable}. Following Gao et al.\ \citeyearpar{Ref24001}, where the market price of risk of the traded asset is fixed and only the incomplete-market price of risk is optimised, the market-price-of-risk vector optimised in Section~\ref{refsection3} is
\[
\theta=(\lambda^\prime,\kappa)\qquad\big(\text{equivalently } \vartheta=(\eta_f,\kappa)\big).
\]
\end{remark}

\noindent The next section focuses on finding the pricing measure under which the CAT bond contract will be valued.

\section{Valuation framework and pricing algorithm}\label{refsection3}
In the classical asset-pricing framework, the assumption of an arbitrage-free market guarantees the existence of an equivalent martingale measure $\mathbb{Q}$, equivalent to the reference measure $\mathbb{P}$. Under this measure, any discounted price process $\{\tilde{V}(t): 0 \leq t \leq T\}$ of a traded asset is a martingale. Consequently, the current value of a claim paying $V(T)$ at maturity $T$ can be expressed as
\begin{equation}\label{eq:martingale_pricing}
V^{\mathbb{Q}}(t) \;=\; \mathbb{E}^{\mathbb{Q}}\!\Bigl[D(t, T)\,V(T)\;\Big|\;\mathcal{F}_t\Bigr],
\qquad
D(t,T) \;=\; \exp\Bigl\{-\!\!\int_{t}^{T} r_s \,ds\Bigr\}.
\end{equation}
In a complete market, the corresponding price is unique. However, when the market is incomplete, the set $\mathcal{M}$ of all equivalent martingale measures contains infinitely many possible pricing measures, and selecting an appropriate measure for pricing purposes is of central importance. A common approach is to minimize the Kullback--Leibler divergence (relative entropy) between the real-world measure and its equivalent, among the measures that are consistent with the prices of traded securities. The solution to this optimization yields the minimal entropy martingale measure (see, e.g., Dhaene et al.\ \citeyear{Ref1901}).

In this paper, we adopt the pricing algorithm proposed by Gao et al.\ \citeyearpar{Ref24001}, which employs tools from differential geometry to solve, approximately, a similar optimization problem posed in a continuous-time setting over the infinite-dimensional space $\mathcal{M}$. Rather than working directly with the entire set of equivalent martingale measures, the algorithm restricts attention to a finite-dimensional sub-manifold $\mathcal{S}\subset \mathcal{M}$, which in their Brownian setting is obtained from a truncated Wiener--It\^o chaos expansion of the non-hedgeable market price of risk. It then considers a Riemannian metric to derive a geometrically consistent gradient method that finds the parameters characterizing the optimal equivalent measure. We now elaborate on the optimization procedure.

We define the financial entropy $H_V(\mathbb{Q})$, interpreted as the \textit{negative distance} from the real-world measure $\mathbb{P}$, for a strictly convex function $V(\cdot)$ that is the conjugate dual of a utility function $U(\cdot)$:
\[
H_V(\mathbb{Q}) =
\begin{cases}
   -\mathbb{E}^{\mathbb{P}}\left[V\left(\frac{d\mathbb{Q}}{d\mathbb{P}}\right)\right], & \text{if } \mathbb{E}^{\mathbb{P}}\left[\left|V\left(\frac{d\mathbb{Q}}{d\mathbb{P}}\right)\right|\right] < \infty, \\
   -\infty, & \text{otherwise}.
\end{cases}
\]
Solving the following optimization problem over $\mathcal{M}$ is generally intractable due to its infinite-dimensional nature:
\begin{eqnarray}\label{ref1001}
\mathbb{Q}^* = \arg\max_{\mathbb{Q} \in \mathcal{M}} H_V(\mathbb{Q}).
\end{eqnarray}
To overcome this difficulty, we solve a more tractable problem over a finite-dimensional sub-manifold $\mathcal{S} \subset \mathcal{M}$:
\begin{eqnarray}\label{ref1002}
\mathbb{Q}_{\theta^*} = \arg\max_{\mathbb{Q}_\theta \in \mathcal{S}} H_V(\mathbb{Q}_\theta),
\end{eqnarray}
where $\mathcal S$ is a finite-dimensional family of equivalent measures parameterised by a coordinate vector $\theta$; the construction used by Gao et al.\ \citeyearpar{Ref24001} is recalled in Appendix~\ref{app:gln}.
The critical point of problem (\ref{ref1002}) is obtained by ascending in the direction of the Riemannian gradient of the financial entropy functional on $\mathcal{S}$,
\begin{eqnarray}\label{eq:riemgrad}
(\nabla H_V)_i = \sum_{j=1}^{n} g^{ij} \frac{\partial H_V(\mathbb{Q_{\theta}})}{\partial \theta_j}, \quad 1\leq i \leq n,
\end{eqnarray}
where $g^{ij}$ is the $(i,j)$-th entry of the inverse of the Riemannian metric matrix $(g_{ij})$ on $\mathcal{S}$. The metric $g_{ij}$ arises from the second-order partial derivatives of the Bregman distance\footnote{The Bregman distance between two points of $\mathcal{M}_V = \{\mathbb{Q}\in \mathcal{M} : \mathbb{E}^{\mathbb{P}}[|V(\frac{d\mathbb{Q}}{d\mathbb{P}})|]<\infty \}$ is $D_V(\mathbb{Q}_1, \mathbb{Q}_2) = \mathbb{E}^{\mathbb{P}}[\Delta_V(\frac{d\mathbb{Q}_1}{d\mathbb{P}}, \frac{d\mathbb{Q}_2}{d\mathbb{P}})]$, where $\Delta_V(w, x) = V(w) - V(x) - V^{\prime}(x)(w - x)$.} in the Riemannian Metric Theorem of Gao et al.\ \citeyearpar{Ref24001}, which generalizes Euclidean distance to non-Euclidean geometries and measures the divergence between probability measures on $\mathcal{S}$:
\begin{eqnarray}
g_{ij} = \mathbb{E}^{\mathbb{P}}\left[V^{\prime\prime}\left(\frac{d\mathbb{Q_{\theta}}}{d\mathbb{P}}\right)\frac{\partial}{\partial \theta_i}\left(\frac{d\mathbb{Q_{\theta}}}{d\mathbb{P}}\right)\frac{\partial}{\partial \theta_j}\left(\frac{d\mathbb{Q_{\theta}}}{d\mathbb{P}}\right)\right].
\end{eqnarray}
Assuming the exponential utility function $U(x) = -e^{-x}$, the corresponding convex conjugate function\footnote{Other candidates are the log utility function $U(x) = \log x$ and the power hyperbolic absolute risk aversion utility function $U(x)= -qx^p$, where $p<1$, $p\neq 0$, and $\frac{1}{p}+\frac{1}{q}= 1$, for which $V(z) = -\log z$ and $V(z)= \mathrm{sign}(p)z^q$, respectively.} is $V(z) = z\log(z)$ for $z>0$. Consequently, the financial entropy becomes minus the relative entropy,
\begin{eqnarray}
H_V(\mathbb{Q}_{\theta}) = -\mathbb{E}^{\mathbb{P}}\left[\frac{d\mathbb{Q}_{\theta}}{d\mathbb{P}}\log\Big(\frac{d\mathbb{Q}_{\theta}}{d\mathbb{P}}\Big)\right]=-\mathbb{E}^{\mathbb{Q}_{\theta}}\left[\log\Big(\frac{d\mathbb{Q}_{\theta}}{d\mathbb{P}}\Big)\right]=-\mathrm{KL}(\mathbb Q_\theta\|\mathbb P),
\end{eqnarray}
and the Riemannian metric of $\mathcal{S}$ is the Fisher information metric
\begin{eqnarray}\label{eq:fisher}
g_{ij} = \mathbb{E}^{\mathbb{P}}\left[\frac{d\mathbb{Q_{\theta}}}{d\mathbb{P}}\frac{\partial}{\partial \theta_i}\left(\log\frac{d\mathbb{Q_{\theta}}}{d\mathbb{P}}\right)\frac{\partial}{\partial \theta_j}\left(\log \frac{d\mathbb{Q_{\theta}}}{d\mathbb{P}}\right)\right].
\end{eqnarray}

\paragraph{Our manifold.} In our application, $\mathcal S$ is the two-dimensional family of measures $\mathbb Q_\theta$ defined by \eqref{ref51001}, with coordinates $\theta=(\lambda^\prime,\kappa)$ and with $\lambda_r$ fixed as in Remark~\ref{rem:hedge}. This is a parametric jump family rather than a chaos truncation. We therefore use the financial entropy, the Riemannian metric and the gradient method of Gao et al.\ \citeyearpar{Ref24001}, but we do not rely on their results that are specific to the Brownian setting. These are the density of chaos truncations in $\mathcal M$, their existence-and-uniqueness result (which assumes relative risk aversion $R(x)\le1$ for all $x$ and does not cover exponential utility), and their minimal-pricing-model-risk interpretation.

\subsection{Financial entropy of the CAT family and the role of market information}
For our family, both the financial entropy and the Riemannian metric are available in closed form (proofs in Appendix~\ref{app:proofs}).

\begin{proposition}[Closed-form financial entropy]\label{prop:KL}
For $\mathbb Q$ defined by \eqref{ref51001},
\begin{eqnarray}\label{eq:KL}
-H_V(\mathbb Q)&=&\mathrm{KL}(\mathbb Q\|\mathbb P)\nonumber\\
&=&\underbrace{\lambda T\Big[(1+\lambda^\prime)\ln(1+\lambda^\prime)-\lambda^\prime+(1+\lambda^\prime)\frac{\kappa^2}{2}\Big]}_{=:\mathcal K_C(\theta)}
+\underbrace{\frac{\lambda_r^2}{2\sigma_r^2}\int_0^T\mathbb E^{\mathbb Q}[r(s)]\,ds}_{=:\mathcal K_r(\lambda_r)} .
\end{eqnarray}
In the coordinates $\vartheta=(\eta_f,\kappa)$, $\mathcal K_C=\lambda T\big[e^{\eta_f}(\eta_f-1+\kappa^2/2)+1\big]$.
\end{proposition}

\begin{proposition}[Entropy without market information]\label{prop:degenerate}
The maximiser of $H_V$ over $(\lambda_r,\lambda^\prime,\kappa)$ is $\lambda_r=\lambda^\prime=\kappa=0$, i.e.\ $\mathbb Q^*=\mathbb P$, for which the CAT bond price equals its physical value.
\end{proposition}
\begin{proof}
Both terms of \eqref{eq:KL} are non-negative and vanish only at $\lambda_r=0$ and at $\lambda^\prime=\kappa=0$, respectively.
\end{proof}
\noindent This is in line with Dhaene et al.\ \citeyearpar{Ref1901}, who show in a one-period setting that, when no actuarial risks are traded, the minimal entropy martingale measure leaves the distribution of the actuarial risks unchanged.

\noindent In Gao et al.\ \citeyearpar{Ref24001} this degeneracy does not arise, because the market price of risk of the traded asset is fixed by no-arbitrage and $\mathbb P\notin\mathcal M$. In our setting the corresponding information is (i) the Treasury curve, which fixes $\lambda_r$ (Remark~\ref{rem:hedge}), and (ii) the observed CAT bond spreads, which must restrict the catastrophe premia $\theta$. Let $V^{\mathrm{model}}_{t_k}(\theta)$ denote the model price at the quotation date $t_k$, $k=0,\dots,K$, of Assumption~\ref{ass:traded} (Section~\ref{sec:payoff}) and $V^{\mathrm{mkt}}_{t_k}$ the market value; in the Li et al.\ \citeyearpar{Ref989800} set-up used here, $V^{\mathrm{mkt}}_{t_k}$ is par (Section~\ref{ref54}). \paragraph{Pooled criterion.} To select a single measure for the whole life of the bond, we use the \emph{market-consistent entropy criterion}
\begin{eqnarray}\label{eq:J}
\theta^*_\alpha=\arg\max_{\theta}\ \mathcal J_\alpha(\theta),\qquad
\mathcal J_\alpha(\theta):=H_V(\mathbb Q_\theta)-\frac{1}{2\alpha}\sum_{k=0}^{K}\Big(V^{\mathrm{model}}_{t_k}(\theta)-V^{\mathrm{mkt}}_{t_k}\Big)^2,
\end{eqnarray}
with $\alpha>0$. Problem \eqref{eq:J} is the entropy-regularised calibration of Cont and Tankov \citeyearpar{RefCT04}, with the physical measure as prior: $\alpha$ governs the trade-off between fitting the observed prices and staying close to $\mathbb P$. As $\alpha\downarrow0$ the criterion reduces to non-linear least squares, which by Proposition~\ref{prop:ident} is nearly flat along a curve of parameters; the entropy term selects among them. We choose $\alpha$ by the Morozov discrepancy principle, as in Cont and Tankov (\citeyear{RefCT04}, Section 4.3): the largest $\alpha$ for which the root-mean-square pricing error does not exceed the observation noise level. We take the noise level to be $0.25$ per $100$ of face value, half of a typical dealer bid--ask range of half a point. If the model cannot fit the quotes to this accuracy for any $\alpha$, the observation noise is not the binding error, and we use the same principle with the noise level estimated from the best fit, $\hat\sigma^2=\mathrm{SSE}_{\min}/(n-2)$: we take the largest $\alpha$ whose sum of squared errors does not exceed $n\hat\sigma^2$, with $n$ the number of quotes.

\paragraph{Date-wise criterion.} To track how the market priced catastrophe risk over time, we also select, at each quarter $t_k$ separately, the maximum-entropy measure that reproduces that quarter's price exactly:
\begin{eqnarray}\label{eq:DW}
\theta^*_k=\arg\max_{\theta\in\Theta}\ H_V(\mathbb Q_\theta)\qquad\text{subject to}\qquad V^{\mathrm{model}}_{t_k}(\theta)=V^{\mathrm{mkt}}_{t_k},\qquad k=0,\dots,K,
\end{eqnarray}
where $\Theta$ is a compact box of economically admissible values (e.g.\ $|\eta_f|,|\kappa|\le 5$). Each problem has one equality constraint and two unknowns, so its feasible set is (generically) a curve in $\Theta$, and \eqref{eq:DW} selects the point of this curve that is closest to $\mathbb P$. Only information available at $t_k$ is used, so the resulting series $\{\theta^*_k\}$ contains no look-ahead. By Proposition~\ref{prop:ident} below, the implied trigger intensity $\ell(\theta^*_k)$ is essentially determined by the price, while the entropy determines the split between $\lambda^{\prime}$ and $\kappa$. The solution $\theta^*_k$ is the measure in the family that is closest to $\mathbb P$ in relative entropy among those that reproduce the observed price, i.e.\ an in-family version of the minimal-entropy consistent martingale measure of Kallsen \citeyearpar{RefKallsen01}; see also Cont and Tankov (\citeyear{RefCT04}, Section 3.3.1).

\begin{proposition}[Existence for the date-wise problem]\label{prop:DWexist}
Suppose $\theta\mapsto V^{\mathrm{model}}_{t_k}(\theta)$ is continuous on $\Theta$ and the feasible set $\mathcal F_k=\{\theta\in\Theta:V^{\mathrm{model}}_{t_k}(\theta)=V^{\mathrm{mkt}}_{t_k}\}$ is non-empty. Then \eqref{eq:DW} has a solution. In the per-occurrence approximation of Proposition~\ref{prop:ident}, every interior solution satisfies the closed-form condition \eqref{eq:foc}, with $\ell^*$ the trigger intensity implied by the price at $t_k$.
\end{proposition}
\begin{proof}
$\mathcal F_k$ is a closed subset of the compact set $\Theta$ and $H_V(\mathbb Q_\theta)=-\mathcal K_C(\theta)-\mathcal K_r$ is continuous in $\theta$ (Proposition~\ref{prop:KL}), so the maximum is attained. The second statement is shown in Appendix~\ref{app:proofs}.
\end{proof}

\noindent A non-empty $\mathcal F_k$ requires that the observed price lies between the model price with no trigger risk and the price with immediate trigger. Whether this holds depends on the data and on how quoted spreads are translated into prices; we check it at every quotation date in Section~\ref{sec:datewise}. Uniqueness is not proved; in two dimensions it is verified by evaluating $\mathcal K_C$ along the feasible curve.

\subsection{Pricing algorithms}
Building upon the preceding discussion, the pricing algorithm for any discounted contingent claim (denoted as DC) in the pooled analysis is the Riemannian gradient method of Gao et al.\ \citeyearpar{Ref24001} applied to the objective $\mathcal J_\alpha$:
\begin{algorithm}[H]
\caption{Riemannian geometric pricing algorithm (market-consistent)}\label{alg:1}
\begin{algorithmic}[1]
\Require Initial guess $\theta^{(0)}$ (na\"ive least squares, Section~\ref{ref54}), step sizes $\{\tau_k\}$, tolerance $\epsilon > 0$, $\alpha>0$
\Ensure Optimal parameter $\theta^*$ and $\mathbb{E}^{\mathbb{Q}_{\theta^*}}[\text{DC}]$ as the price
\State Set iteration counter $k \gets 0$
\Repeat
    \State Compute $\frac{\partial}{\partial \theta_i}\mathcal J_\alpha(\theta^{(k)})$ \Comment{entropy part in closed form; price part by finite differences}
    \State Compute metric tensor $g_{ij}(\theta^{(k)})$ \Comment{closed form, Proposition~\ref{prop:metric}}
    \State Invert $g_{ij}$ to obtain $g^{ij}$
    \State Compute the Riemannian gradient $(\nabla \mathcal J_\alpha)_i = \sum_{j} g^{ij}\, \partial \mathcal J_\alpha/\partial \theta_j$
    \State Update $\theta^{(k+1)} = \theta^{(k)} + \tau_k \nabla \mathcal J_\alpha$ \Comment{$\tau_k$ by backtracking line search}
    \State Set $k \gets k + 1$
\Until{$\big\|\nabla \mathcal J_\alpha(\theta^{(k)})\big\|_{g} < \epsilon$}
\State \Return $\theta^* = \theta^{(k)}$ and $\mathbb{E}^{\mathbb{Q}_{\theta^*}}[\text{DC}]$
\end{algorithmic}
\end{algorithm}
\noindent Convergence is local, as for any gradient method. When the problem has a unique maximiser (which can be checked on a grid in two dimensions), different initialisations can only differ in computational cost, not in the final answer. We therefore start Algorithm~\ref{alg:1} from a single initialisation, the na\"ive least-squares estimate, and check its output against a grid search over the two-dimensional parameter space (Section~\ref{ref54}).

For the two-parameter family used here, the entropy and the metric are available in closed form, and the date-wise problem reduces to a one-dimensional search along the curve $\{\ell(\theta)=\ell^*\}$, which we use as an independent check. The Riemannian (natural-gradient) formulation is nevertheless useful for three reasons. First, it is invariant to the parameterisation: the coordinates $\lambda^\prime$ and $\eta_f=\ln(1+\lambda^\prime)$ give the same steps, whereas a Euclidean gradient depends on the chosen coordinates and on the units of $\lambda^\prime$ and $\kappa$. Second, it removes the scale of the problem: the Euclidean gradient of $\mathcal K_C$ is proportional to $\lambda T$, which is about $0.09$ in our application, whereas the natural gradient (Proposition~\ref{prop:metric}) does not depend on $\lambda$ or $T$, so that no tuning of step sizes is needed. Third, the algorithms do not rely on the closed form of the constraint curve. They apply unchanged to richer families of measures (more parameters, time-varying premia, several bonds calibrated jointly, or utilities other than the exponential), where no reduction to a one-dimensional search is available.

\noindent For the date-wise problem \eqref{eq:DW}, the gradient step must keep the constraint satisfied. We use the Riemannian projected-gradient method with a restoration step (Algorithm~\ref{alg:2}): the Riemannian gradient of the entropy is projected onto the tangent space of the constraint curve, and a Gauss--Newton step along the metric-gradient of the constraint restores feasibility. This is a natural-gradient method in the sense of Amari \citeyearpar{RefAmari98} on the constraint curve. Its output is validated by a direct one-dimensional search along the feasible curve.

\begin{algorithm}[H]
\caption{Date-wise Riemannian algorithm with constraint restoration (quarter $t_k$)}\label{alg:2}
\begin{algorithmic}[1]
\Require Initial guess $\theta^{(0)}$ (e.g.\ $\kappa=0$ and $\lambda^\prime$ solving the constraint), step sizes $\{\tau_n\}$, tolerances $\epsilon_H,\epsilon_c$
\For{$n=0,1,2,\dots$}
  \State $c\gets V^{\mathrm{model}}_{t_k}(\theta^{(n)})-V^{\mathrm{mkt}}_{t_k}$,\quad $a\gets\nabla_\theta V^{\mathrm{model}}_{t_k}(\theta^{(n)})$
  \State $v\gets g^{-1}\nabla_\theta H_V(\mathbb Q_{\theta^{(n)}})$,\quad $w\gets g^{-1}a$ \Comment{closed-form metric, Proposition~\ref{prop:metric}}
  \State $v_T\gets v-\dfrac{a^\top v}{a^\top w}\,w$ \Comment{projection onto the tangent space of the constraint}
  \State $\delta\gets-\dfrac{c}{a^\top w}\,w$ \Comment{restoration towards the constraint}
  \If{$\|v_T\|_g<\epsilon_H$ \textbf{and} $|c|<\epsilon_c$} \textbf{break} \EndIf
  \State $\theta^{(n+1)}\gets\theta^{(n)}+\tau_n v_T+\delta$
\EndFor
\State \Return $\theta^*_k=\theta^{(n)}$, the implied intensity $\ell(\theta^*_k)$ and $\mathbb{E}^{\mathbb{Q}_{\theta^*_k}}[\text{DC}]$
\end{algorithmic}
\end{algorithm}

\subsection{CAT bond payoff structure and its value}\label{sec:payoff}
We now apply the pricing algorithm to a CAT bond contract. Suppose that $\{t_1 , t_2, \cdots, t_m = T\}$ are the coupon-paying dates with constant spacing $\Delta$ (one, three or six months, i.e.\ $\Delta=1/12$, $3/12$ or $6/12$ year). Let $C\geqslant 0$ be the constant annual coupon rate, $\zeta\in[0,1)$ the fraction of principal paid to the investor if the bond is triggered, and
\[
\tau = \inf\{0\leqslant t \leqslant T : L_t \geqslant D\}
\]
the first time the aggregate loss reaches the threshold level $D$ (aggregate trigger). Because $L$ is non-decreasing, $\{\tau>t\}=\{L_t<D\}$.

\noindent Once the contract starts, the holder of the CAT bond receives the constant amount $C\Delta N$ at each coupon date $t_i$ at which the bond has not been triggered. At maturity $T$, the issuer pays back the principal $N$ provided that the bond has not been triggered. Whenever $L_{\tau}\geqslant D$ for some $\tau \in (t_{i-1}, t_i]$, the contract is terminated and the investor receives $\zeta N$ \emph{once}, at $t_i$. In the application of Section~\ref{ref54}, the coupon rate is reset at each quotation date to the observed market spread plus the risk-free par yield, and is held constant over the remaining life of the bond valued at that date. Mathematically,
\begin{eqnarray*}
Pay_{CAT}(t_i) = \left\{ \begin{array}{ll }
C\Delta N\, I_{\{\tau > t_i\}} + \zeta N\, I_{\{t_{i-1}<\tau \leqslant t_i\}}, & i=1,\dots,m-1,\\[3pt]
\left[C\Delta N +N\right] I_{\{\tau > T\}} + \zeta N\, I_{\{t_{m-1}<\tau \leqslant T\}}, & i=m.
\end{array}\right.
\end{eqnarray*}

\noindent At a valuation date $t_k$ at which the bond is alive, investors know that the bond has not been triggered but do not observe the modelled loss index itself. We therefore condition on $\{\tau>t_k\}$ and define the conditional survival probability
\begin{eqnarray}\label{eq:surv}
S^{\mathbb Q}_{k}(t):=\mathbb Q(\tau>t\mid\tau>t_k)=\frac{\mathbb Q(L_t<D)}{\mathbb Q(L_{t_k}<D)},\qquad
\mathbb Q(L_t<D)=\sum_{n\ge0}e^{-\hat\lambda t}\frac{(\hat\lambda t)^n}{n!}\,\hat F^{*n}(D),
\end{eqnarray}
where $\hat F^{*n}$ is the $n$-fold convolution of \eqref{eq:wang}. It can be evaluated by Panjer recursion (Panjer \citeyear{RefPanjer81}), by the fast Fourier transform after discretising the claim-size distribution (Mildenhall \citeyear{RefMildenhall25}), or by Monte Carlo simulation. Since only probabilities below the threshold $D$ are needed, truncating the heavy-tailed claim-size distribution at or above $D$ does not affect $\mathbb Q(L_t<D)$ (Mildenhall \citeyear{RefMildenhall25}, Section 2.4.3); in our application the FFT values agree with Monte Carlo estimates (Appendix~\ref{app:mc}). Based on the payoff structure and the valuation formula (\ref{eq:martingale_pricing}), and using the independence of $W$ and $L$ under $\mathbb Q$, the fair price at $t_k$ is
\begin{eqnarray}\label{ref18}
\begin{gathered}[b]
V_{t_k}^{model}(C;\theta) = \mathbb{E}^{\mathbb{Q}_\theta}\left[\sum_{j=k+1}^{m}D(t_k, t_j)\,Pay_{CAT}(t_j)\;\Big|\;\mathcal F^W_{t_k},\ \tau>t_k\right]\\
= \sum_{j=k+1}^{m} \mathcal P_{k,j}\Big(C\Delta N\, S^{\mathbb Q}_k(t_j) + \zeta N \big[S^{\mathbb Q}_k(t_{j-1})-S^{\mathbb Q}_k(t_j)\big]\Big)\\
+\mathcal P_{k,m}\,N\,S^{\mathbb Q}_k(T),
\end{gathered}
\end{eqnarray}
where $S^{\mathbb Q}_k(t_k)=1$ and $\mathcal P_{k,j}$ is the time-$t_k$ price of a default-free zero-coupon bond maturing at $t_j$. In the main analysis $\mathcal P_{k,j}=P^{\mathrm{mkt}}(t_k,t_j)$, the discount factor bootstrapped from the Treasury curve observed at $t_k$ (Remark~\ref{rem:hedge}). In the CIR robustness check, $\mathcal P_{k,j}=P(t_k, t_j , r(t_k) , \theta^*_{r} , m^*_{r}, \sigma_r )$ from Appendix A, relation (\ref{ref101}), with $r(t_k)$ the observed short rate and $\lambda_r$ calibrated to the same curve. Either way, the model price depends on the market-price-of-risk vector $\theta$ only through the survival probabilities.

\subsection{Riemannian metric of the CAT family}
Using historical data, we first estimate the physical parameters governing the loss distribution, frequency distribution and interest-rate dynamics. The market-price-of-risk vector $\theta = (\lambda', \kappa)$ is then determined by Algorithm~\ref{alg:1}, for which the Riemannian metric $g_{ij}$ is derived as follows. Define the score functions
\begin{eqnarray}
    G_1(\theta) &=& \frac{\partial}{\partial\lambda^\prime}\log\frac{d\mathbb Q_\theta}{d\mathbb P}=-\lambda T + \frac{N(T)}{1+ \lambda^\prime},\\
    G_2(\theta) &=& \frac{\partial}{\partial\kappa}\log\frac{d\mathbb Q_\theta}{d\mathbb P}=\sum_{i=1}^{N(T)} \frac{\partial}{\partial\kappa}[\log g(X_i)] = \sum_{i=1}^{N(T)}\big[\Phi^{-1}(F(X_i))-\kappa\big].
\end{eqnarray}
The components of the Riemannian metric are $g_{11} = \mathbb{E}^{\mathbb{P}}[\frac{d\mathbb{Q_{\theta}}}{d\mathbb{P}}G_1(\theta)^2]$, $g_{22} = \mathbb{E}^{\mathbb{P}}[\frac{d\mathbb{Q_{\theta}}}{d\mathbb{P}}G_2(\theta)^2]$ and $g_{12} = g_{21} = \mathbb{E}^{\mathbb{P}}[\frac{d\mathbb{Q_{\theta}}}{d\mathbb{P}}G_1(\theta)G_2(\theta)]$, and they can be computed explicitly.

\begin{proposition}[Closed-form Riemannian metric]\label{prop:metric}
In the coordinates $\theta=(\lambda^\prime,\kappa)$ and $\vartheta=(\eta_f,\kappa)$, respectively,
\begin{eqnarray}\label{eq:metric}
(g_{ij})=\lambda T\begin{pmatrix}\frac{1}{1+\lambda^\prime}&0\\[2pt]0&1+\lambda^\prime\end{pmatrix},
\qquad (g_{ij})_{\vartheta}=\lambda T\,e^{\eta_f}\,I_2 .
\end{eqnarray}
The entropy gradient is $\nabla_\vartheta H_V=-\lambda T e^{\eta_f}\big(\eta_f+\tfrac{\kappa^2}{2},\ \kappa\big)^\top$, so the Riemannian gradient of the entropy part is simply $-\big(\eta_f+\tfrac{\kappa^2}{2},\ \kappa\big)^\top$.
\end{proposition}
\noindent Thus no Monte Carlo is needed for the entropy part of Algorithm~\ref{alg:1}; Monte Carlo (or recursion) is needed only for the prices \eqref{ref18}. The closed forms were checked against Monte Carlo estimates with $\XmcN\times10^6$ simulated paths (Table~\ref{tab:mc} in Appendix~\ref{app:mc}); all estimates lie within $\XmcZ$ standard errors of the closed forms.

\begin{remark}[Structure of the metric]\label{rem:metric}
For a market-price-of-risk vector with $d$ components, the metric $(g_{ij})$ is a $d\times d$ matrix whose entries are the covariances, under $\mathbb Q_\theta$, of the score functions. In our family the off-diagonal entries vanish because, under $\mathbb Q_\theta$, the number of events and the claim sizes are independent, so that the scores $G_1$ and $G_2$ are uncorrelated. Additional parameters enlarge the matrix: if the interest-rate premium $\lambda_r$ were included as a coordinate, it would add the diagonal entry $\sigma_r^{-2}\int_0^T\mathbb E^{\mathbb Q}[r(s)]\,ds$, again with zero cross terms. Under dependence between risk drivers, such as between claim sizes and inter-arrival times (Shao et al.\ \citeyear{Ref38}), the scores are in general correlated and the off-diagonal entries are non-zero; the Riemannian gradient and Algorithms~\ref{alg:1}--\ref{alg:2} apply unchanged, with the metric estimated by Monte Carlo averages of products of scores when no closed form is available.
\end{remark}

\subsection{What the spreads can and cannot identify}
\begin{figure}[htbp]
\centering
\includegraphics[width=\linewidth]{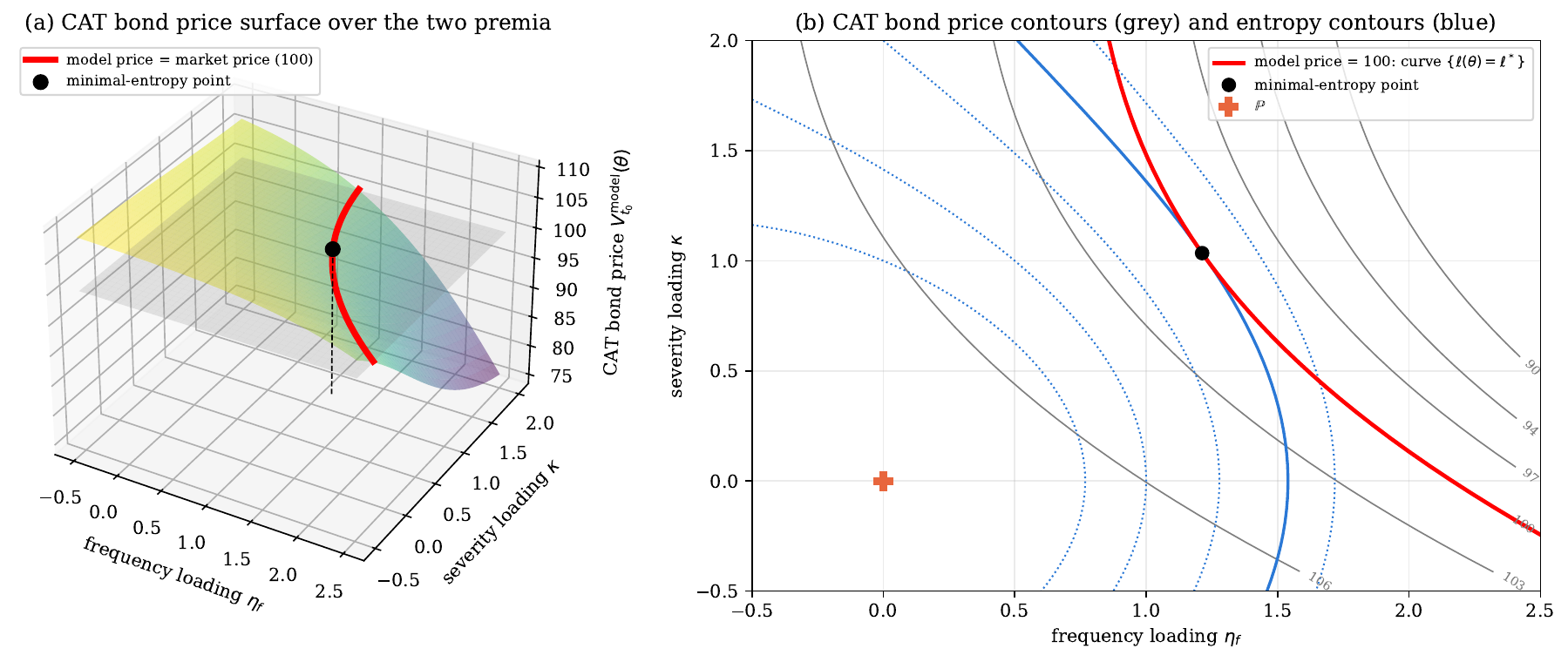}
\caption{Geometry of the date-wise problem at the issue date (illustration based on the per-occurrence approximation with flat discounting). (a) CAT bond price $V^{\mathrm{model}}_{t_0}(\theta)$ as a function of the frequency and severity loadings; the grey plane is the market price 100, and the red curve, where the price surface meets it, contains all premia that reproduce the market price. (b) The same seen from above: price contours (grey), entropy contours (blue) and the physical measure $\mathbb P$; the minimal-entropy point is where the smallest entropy contour touches the red curve.}
\label{fig:geometry}
\end{figure}

\begin{proposition}[Weak identification]\label{prop:ident}
Let $\ell(\theta):=\hat\lambda\,\bar{\hat F}(D)=\lambda(1+\lambda^\prime)\,\Phi\big(\kappa-\Phi^{-1}(F(D))\big)$ be the risk-neutral intensity of events that on their own exceed $D$. For $s=\hat\lambda t$,
\begin{eqnarray}\label{eq:expansion}
\mathbb Q(L_t<D)=e^{-\ell(\theta)t}-\frac{s^2}{2}e^{-s}\Big(\hat F(D)^2-\hat F^{*2}(D)\Big)+O(s^3).
\end{eqnarray}
Hence, up to terms of order $(\hat\lambda t)^2$, all prices \eqref{ref18} depend on $\theta$ only through $\ell(\theta)$. For a per-occurrence trigger the dependence is exactly through $\ell(\theta)$, for every $t$.
\end{proposition}
\noindent At the pooled optimum $\theta^*$ (Section~\ref{ref54}), $\hat\lambda T=\XlamHatT$, so the correction term in \eqref{eq:expansion} is small. The approximation is local: it requires $\hat\lambda T$ to be small, which holds near $\theta^*$ but not at the na\"ive least-squares estimate ($\lambda^\prime\approx\XlpLS$, where $\hat\lambda T=\XlamHatTls$), nor at the boundary $\lambda^\prime=e^5-1$ of $\Theta$ ($\hat\lambda T=\XlamHatTbound$). In the region of economically plausible premia, the observed spreads therefore pin down the risk-neutral trigger intensity well, but they carry almost no information on how it splits into frequency ($\lambda^\prime$) and severity ($\kappa$) loadings: all $\theta$ on the curve $\{\ell(\theta)=\ell^*\}$ price the bond (almost) identically. The sum-of-squares surface of the na\"ive calibration is therefore nearly flat along this curve, and its minimum is determined by the second-order term in \eqref{eq:expansion}, i.e.\ by the accumulation of many sub-threshold losses at large $\hat\lambda T$. This is where the entropy term of \eqref{eq:J} does the work: it selects the point on the curve closest to $\mathbb P$. In the per-occurrence case the selected point satisfies the closed-form condition
\begin{eqnarray}\label{eq:foc}
\kappa^*=M\big(\kappa^*-u_D\big)\Big(\eta_f^*+\frac{\kappa^{*2}}{2}\Big),\qquad \eta_f^*=\ln\frac{\ell^*}{\lambda}-\ln\Phi(\kappa^*-u_D),
\end{eqnarray}
where $u_D=\Phi^{-1}(F(D))$ and $M(x)=\phi(x)/\Phi(x)$; the derivation is given in Appendix~\ref{app:proofs}. This provides an independent check of the output of Algorithm~\ref{alg:2}. Figure~\ref{fig:geometry} illustrates this geometry.

\begin{corollary}[Minimal-entropy path]\label{cor:path}
Under a per-occurrence trigger, suppose that the minimiser of $\mathcal K_C$ on the curve $\{\ell(\theta)=\ell^*\}$ is unique for every $\ell^*>0$, and denote it by $\theta^{\mathrm{ME}}(\ell^*)$. Then $\theta^{\mathrm{ME}}$ does not depend on the quotation date, on the horizon $T$, or on the contract terms other than $u_D$, and the date-wise solutions satisfy $\theta^*_k=\theta^{\mathrm{ME}}(\ell^*_k)$. All date-wise solutions therefore lie on the curve $\{\theta^{\mathrm{ME}}(\ell):\ell>0\}$, which passes through $\theta=0$ at $\ell=\lambda\bar F(D)$, and the time variation of the selected split is a deterministic function of the time variation of $\ell^*_k$. Under an aggregate trigger the same holds up to the correction term in \eqref{eq:expansion}.
\end{corollary}
\begin{proof}
$\mathcal K_C$ is proportional to $\lambda T$, and by Proposition~\ref{prop:ident} the constraint depends on the date only through $\ell^*_k$.
\end{proof}

\section{Valuation under physical measure $\mathbb{P}$}\label{ref009}
For illustrative purposes, we derive the CAT bond value under the physical measure $\mathbb{P}$ in order to compare it with the value obtained under the optimal pricing measure $\mathbb{Q}_{\theta^*}$ determined by our pricing algorithm. To this end, we apply a Feynman--Kac argument (see, e.g., Gan et al.\ \citeyear{Ref24}) to find the expectation of the discount factor under $\mathbb{P}$ rather than $\mathbb{Q}$. This is needed when calculating the expectation of the CAT bond payoff under $\mathbb{P}$. The resulting formula can be cross-checked against the zero-coupon bond formula of Appendix A evaluated at the physical parameters. The proof is given in Appendix B.

\begin{lemma}\label{ref85}
Let $\{r(t): 0\leqslant t \leqslant T\}$ satisfy (\ref{ref3}) and let $u,w\ge0$. Then
\begin{eqnarray}\label{ref87}
g(T, r_0):=\mathbb{E}^{\mathbb{P}}\left[\exp\Big\{-w\int_{0}^{T} r(s)ds - u\,r(T)\Big\}\,\Big|\,r(0)= r_0\right]=\exp\big\{\mathcal A(T)-\mathcal B(T)\,r_0\big\},
\end{eqnarray}
where, with $\varrho=\sqrt{\bar\theta_r^2+2\sigma_r^2w}$, $b_\pm=(-\bar\theta_r\pm\varrho)/\sigma_r^2$ and $c=(u-b_+)/(u-b_-)$,
\begin{eqnarray}
\mathcal B(T)=\frac{b_+-c\,e^{-\varrho T}b_-}{1-c\,e^{-\varrho T}},\qquad
\mathcal A(T)=-\bar\theta_r\bar m_r\Big[b_+T+\frac{2}{\sigma_r^2}\ln\frac{1-c\,e^{-\varrho T}}{1-c}\Big].
\end{eqnarray}
For $u=0$ and $w=1$, $g(T,r_0)$ coincides with the zero-coupon bond formula (\ref{ref101}) evaluated at $(\bar\theta_r,\bar m_r,\sigma_r)$.
\end{lemma}
\begin{corollary}\label{ref88}
By the Markov property, $\mathbb{E}^{\mathbb{P}}\big[\exp\{-w\int_{t}^{T} r(s)ds - u r(T)\}\,\big|\,r(t)\big] = g(T-t, r(t))$.
\end{corollary}

\noindent The CAT bond value at $t_k$ under the physical measure $\mathbb{P}$ is then
\begin{eqnarray}\label{ref10912}
V_{t_k}^{\mathbb P}(C) &=& \sum_{j=k+1}^{m} g(t_j-t_k, r(t_k))\Big(C\Delta N\, S^{\mathbb P}_k(t_j) + \zeta N\big[S^{\mathbb P}_k(t_{j-1})-S^{\mathbb P}_k(t_j)\big]\Big)\nonumber\\
&&+\,g(T-t_k, r(t_k))\,N\,S^{\mathbb P}_k(T),
\end{eqnarray}
where $g$ is given by (\ref{ref87}) with $u=0$ and $w=1$, and $S^{\mathbb P}_k$ is (\ref{eq:surv}) with $(\hat\lambda,\hat F)$ replaced by $(\lambda,F)$. We also report the \emph{actuarial benchmark} $V^{\mathrm{act}}_{t_k}$, which is \eqref{ref18} with the market discount factors $\mathcal P_{k,j}=P^{\mathrm{mkt}}(t_k,t_j)$ and the physical survival probabilities $S^{\mathbb P}_k$. It differs from the model price only through the catastrophe premia. Its difference from the market value,
\begin{eqnarray}\label{eq:premium}
\Pi_{t_k}=V^{\mathrm{act}}_{t_k}-V^{\mathrm{mkt}}_{t_k},
\end{eqnarray}
measures the catastrophe risk premium in price units. The actuarial value is the zero-risk-premium value of Cox and Pedersen \citeyearpar{Ref11}: catastrophe risk is valued with its physical probabilities and interest-rate risk with the market term structure. Accordingly, $\Pi_{t_k}$ is the price-unit counterpart of the excess of the spread over the expected loss used in empirical studies of CAT bond spreads. Since CAT bond spreads are documented to exceed expected losses (e.g.\ Braun \citeyear{Ref9000}), we expect $\Pi_{t_k}>0$, i.e.\ a market price below the actuarial value.

\section{Numerical analysis}\label{ref54}
In this paper, we examine the pricing of a hypothetical catastrophe (CAT) bond with a modelled loss trigger, designed to cover earthquake-related losses. To contextualize the pricing framework, consider an insurance company that provides coverage against earthquake-induced damages. To mitigate its risk exposure, the insurer transfers all or a portion of its liability to a reinsurance company in exchange for a premium. In turn, to manage its own default risk, raise capital and improve diversification, the reinsurance company transfers a portion of its assumed risk to the capital markets by issuing the CAT bond through a special purpose vehicle (SPV). The proceeds from the sale of the CAT bond are fully collateralized in a secure trust account. In the event of a trigger, defined by a modelled loss threshold, a portion of the principal is forfeited, in accordance with the contractual terms agreed upon between the sponsor and the investors.

Risk management techniques can be employed to design a layered compensation system for optimal risk allocation among stakeholders. One reasonable approach is to define the bounds of each layer by quantiles (Value-at-Risk) of the annual aggregate loss. For example, the insurance company retains losses in $[0, \mathrm{VaR}_{0.90})$, the reinsurance company assumes $[\mathrm{VaR}_{0.90}, \mathrm{VaR}_{0.95})$, and the losses above the attachment point $D$ of the third layer are transferred to the capital market through the CAT bond. Since the primary focus of this paper is on pricing, we assume that the risk allocation structure is already established. Our objective is to determine the fair price of the hypothetical CAT bond, assuming it is issued in response to losses that exceed the upper bound of the second risk layer.

To initiate the pricing process, we estimate the physical parameters of the models governing collective loss risk and interest rate risk using historical data. The CIR parameters reported in Safarveisi et al.\ \citeyearpar{Ref8889} ($\bar\theta_r=0.254$, $\bar m_r=0.011$, $\sigma_r=0.074$) were estimated on 1990--2022 data, which extend beyond the valuation dates. To avoid look-ahead we re-estimate the model by exact-transition maximum likelihood (non-central $\chi^2$ transition density) on $\XcirN$ month-end U.S.\ three-month Treasury bill yields from January 1990 to August 2014, with zero yields floored at one basis point. This gives $\bar\theta_r=\XcirK$, $\bar m_r=\XcirM$ and $\sigma_r=\XcirS$; the Feller condition is mildly violated ($2\bar\theta_r\bar m_r=\XcirFL$ versus $\sigma_r^2=\XcirFR$), reflecting the long period of near-zero rates. These parameters enter only the physical benchmark \eqref{ref10912} and the CIR robustness check, since the main analysis discounts with the observed Treasury curve. The collective risk model is a compound Poisson process with generalized Pareto (GPD) severities\footnote{A GPD with shape $a\in \mathbb{R}$ and scale $c>0$ has density $\frac{1}{c}(1+ \frac{ax}{c})^{-(1+ \frac{1}{a})}$.}, with parameters estimated as in Safarveisi et al.\ \citeyearpar{Ref8889}. These were estimated on US earthquake losses from 1906 to 2005, before the valuation period. The losses are expressed in 2020 dollars; since the threshold $D$ is chosen to match the bond's attachment probability, this monetary scaling does not affect any of the results below.

\begin{table}[!ht]
\centering
\caption{Parameter values for the interest-rate model and the collective risk model}
\label{ref20}
\begin{tabular}{l l r}
\hline
Interest-rate process parameters under measure $\mathbb{P}$&\multicolumn{2}{r}{Values}\\
\hline
\multicolumn{2}{l}{$r_0$ \quad Initial instantaneous interest rate}& $\Xrzero\%$ \\
\multicolumn{2}{l}{$\bar{\theta}_r$ \quad Speed of mean reversion} & $\XcirK$ \\
\multicolumn{2}{l}{$\bar{m}_r$\quad Long-run mean} & $\XcirM$ \\
\multicolumn{2}{l}{$\sigma_r$ \quad Instantaneous volatility} & $\XcirS$ \\
\hline
Claim severity parameters under measure $\mathbb{P}$& & \\
\hline
\multicolumn{2}{l}{$a$ \quad Shape parameter} & 2.2 \\
\multicolumn{2}{l}{$c$\quad Scale parameter (USD)} & $2.6 \times 10^{7}$ \\
\hline
Claim frequency parameter under measure $\mathbb{P}$& &  \\
\hline
\multicolumn{2}{l}{$\lambda$ \quad Intensity rate parameter (per year)} & 0.02 \\
\hline
\end{tabular}
\end{table}
\noindent The initial short rate is the observed three-month Treasury bill yield on 30 September 2014, $r_0=\Xrzero\%$. At each later valuation date $t_k$ the observed three-month yield is used as $r(t_k)$, while the CIR parameters are kept at the estimates above. The scale $c$ and the threshold $D$ are in USD.

\subsection{Reference bond and market data}
The market information is taken from the bond issued by \textit{Golden State Re II Ltd.\ (Series 2014-1)}\footnote{\href{https://www.artemis.bm/deal-directory/golden-state-re-ii-ltd-series-2014-1/}{https://www.artemis.bm/deal-directory/golden-state-re-ii-ltd-series-2014-1/}}, sponsored by the California State Compensation Insurance Fund. Table~\ref{tab:contract} summarises its terms as reported in the Lane Financial annual reviews (Lane and Beckwith \citeyear{RefLane}) and in the Artemis deal directory. The bond's modelled-loss trigger covers California earthquake risk for a workers'-compensation portfolio, which makes it a close proxy for our hypothetical earthquake CAT bond in terms of structure and risk profile. Public sources describe the actual bond as per-occurrence, whereas our hypothetical bond uses an aggregate trigger. By Proposition~\ref{prop:ident} the difference in prices is of order $(\hat\lambda T)^2$; numerically, at the pooled optimum ($\hat\lambda T=\XlamHatT$) the per-occurrence and aggregate prices differ by at most $\XoccAgg$ per 100 across the 18 dates (and by $\XoccAggP$ under $\mathbb P$).

\begin{table}[htbp]
\centering\small
\caption{Terms of Golden State Re II Ltd.\ (Series 2014-1) as reported in public sources.}\label{tab:contract}
\begin{tabular}{p{0.38\linewidth}p{0.55\linewidth}}
\toprule
Sponsor & California State Compensation Insurance Fund\\
Covered risk & Workers' compensation losses from California earthquakes (modelled-loss trigger)\\
Size and rating & USD 250 million; S\&P BB+\\
Issue and maturity & September 2014; January 2019 (52 months)\\
Spread at issue & 2.20\% (220 bps) over the collateral return\\
Annual expected loss & 0.25\%\\
Annual attachment / exhaustion probability & 0.49\% / 0.11\% (base catalogue)\\
\bottomrule
\end{tabular}

\smallskip
\begin{minipage}{0.93\linewidth}\footnotesize Sources: Lane and Beckwith \citeyearpar{RefLane}; Artemis deal directory.\end{minipage}
\end{table}

\noindent Following Li et al.\ \citeyearpar{Ref989800}, we assume that the hypothetical bond, had it been issued on 30 September 2014 under the then market conditions, would have traded at the same spreads as the reference bond. As in their study, this assumption is reasonable only if the two bonds share the same risk profile, and we therefore align the threshold $D$ with the reference bond's attachment probability. Li et al.\ align both attachment and exhaustion probabilities because their bond has a layer between the two points. Our bond has a single trigger level, so only the attachment probability is used. With the parameters of Table~\ref{ref20}, $D=6\times10^7$ gives an annual physical trigger probability of about $0.88\%$. Matching $0.49\%$ requires $D=\XD\times10^{8}$~USD, computed from the annual aggregate-loss distribution by FFT. We use the matched $D$ in the main analysis and $D=6\times10^7$ as a robustness check (Section~\ref{sec:robust}).

The market data are obtained from the annual reviews of the catastrophe insurance market published by \textit{Lane Financial L.L.C.}\ (Lane and Beckwith \citeyear{RefLane}), which report quarter-end average secondary-market indications. Our dataset consists of $K+1=18$ observations of market-indicated spreads $s_{t_k}$, recorded at $t_k = \frac{k}{4}$, $k = 0, 1, \dots, K$ with $K=17$ (30 September 2014 to 31 December 2018); see Figure~\ref{fig1}. Apart from two dates, the indications lie between 1.65\% and 2.60\%. The first exception is 31 December 2015, when the indication rises to 4.56\% before falling back to 2.60\% by March 2016. A similar temporary rise at the end of the year is visible for several other low-spread bonds in the same review, and for the sponsor's follow-on bond, Golden State Re II 2018-1, at the end of 2018. This points to a year-end effect in secondary-market indications rather than to a change in the perceived risk of this bond, and we report the results with and without this date. The second is the last quote, of 31 December 2018, eight days before maturity. On the stylised grid the last quote falls at $t_{17}=T$, where the sum in \eqref{ref18} is empty and the model price equals $100$ for every $\theta$; it carries no information on the premia and is excluded from the date-wise analysis (its pooled residual is zero).
\begin{figure}[!htp]
    \centering
    \includegraphics[width=0.8\linewidth]{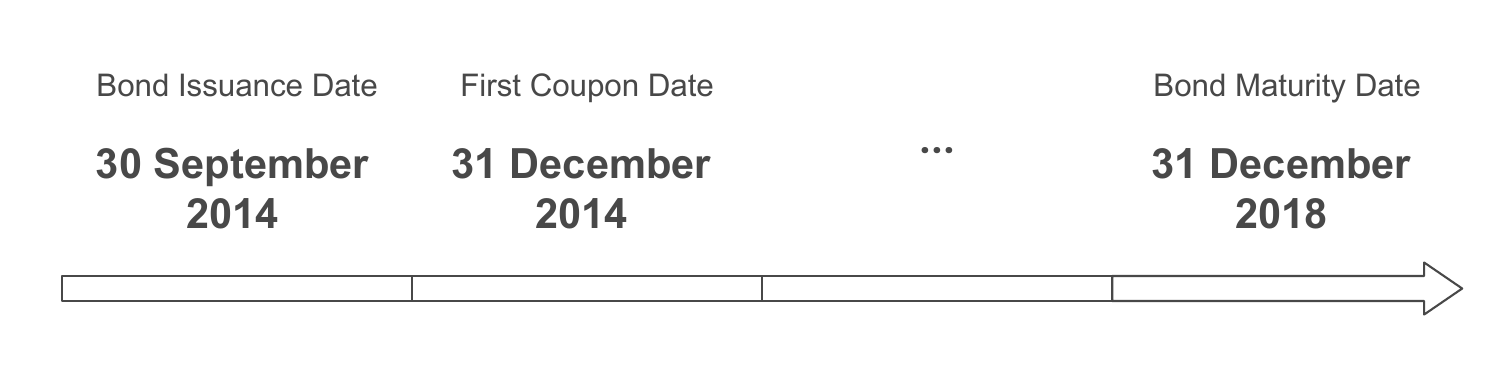}
    \caption{Stylised CAT bond timeline used in the computations.}
    \label{fig2}
\end{figure}
\begin{figure}[!htp]
    \centering
    \includegraphics[width=0.8\linewidth]{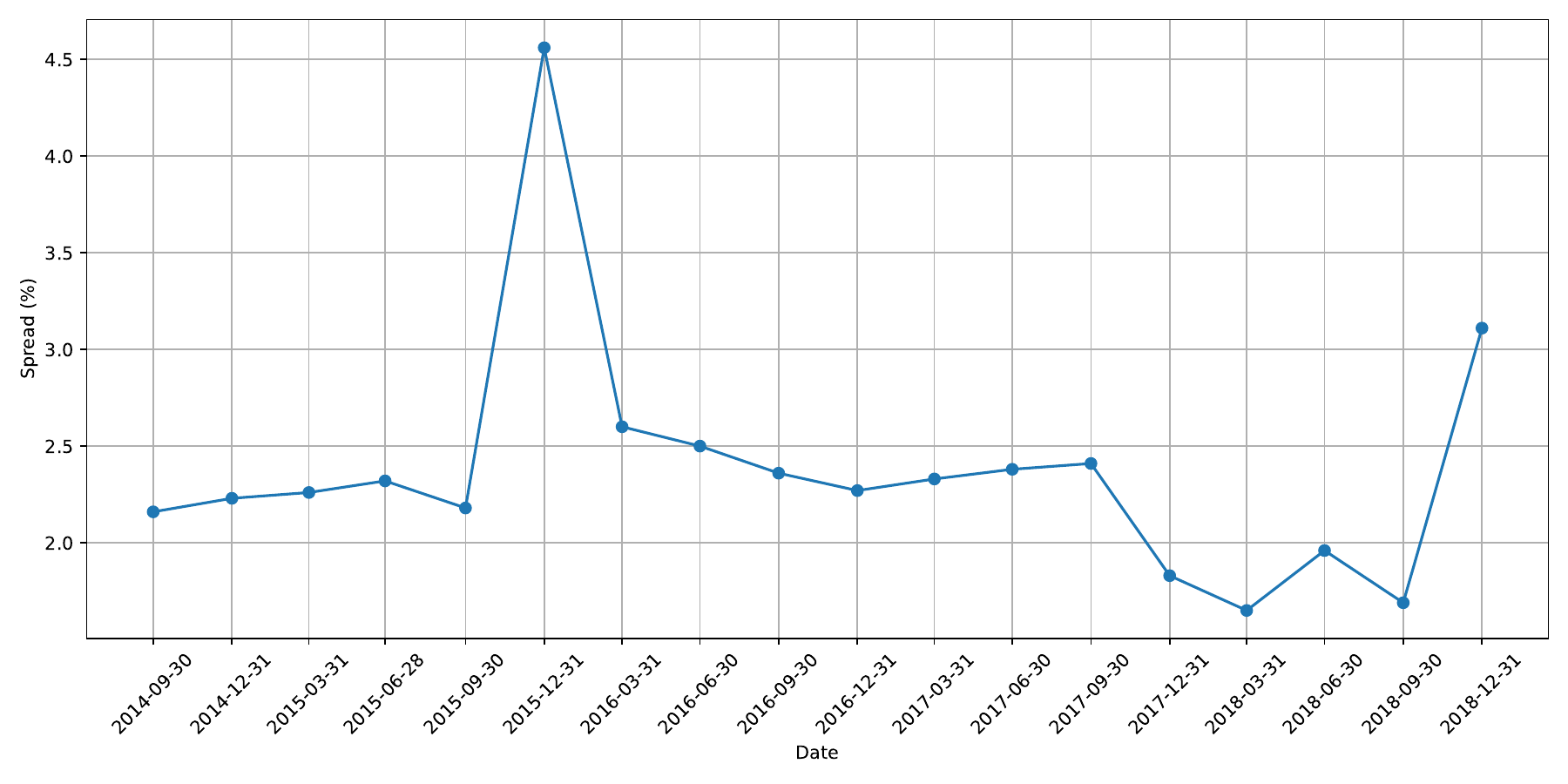}
    \caption{Quarterly market-indicated spreads: Golden State Re II Ltd.\ 2014-1 (Sep 2014 -- Dec 2018). Source: Lane Financial L.L.C.\ annual reviews.}
    \label{fig1}
\end{figure}

Discount factors $P^{\mathrm{mkt}}(t_k,t_j)$ are bootstrapped from the U.S.\ Treasury curve (constant-maturity yields) observed at each quarter-end $t_k$ and interpolated to the coupon dates. We use the U.S.\ Department of the Treasury daily par yield curve (\url{https://home.treasury.gov}), taken on the last business day on or before each $t_k$. Bill yields (1, 3 and 6 months and 1 year) are treated as semi-annually compounded zero yields, note yields (2, 3, 5 and 7 years) as semi-annual par yields, which are bootstrapped; log discount factors are interpolated linearly in maturity (piecewise-flat forward rates). In the base case, each spread results in a model-implied price $V^{\text{model}}_{t_k}(s_{t_k}+y_{t_k};\theta)$ given by relation (\ref{ref18}), with the coupon rate set to the observed spread plus the risk-free par yield $y_{t_k}$ for the remaining term, computed from the same Treasury curve; this approximates the floating coupon (collateral return plus spread) of the actual bond. The model price is required to equal par ($V^{\mathrm{mkt}}_{t_k}=100$ for a face value of 100). The convention of Li et al.\ \citeyearpar{Ref989800}, with the coupon equal to the spread, is reported as a robustness check (Remark~\ref{rem:par}). We adopt a rolling valuation: at each date $t_k$ we fix the coupon at the level $s_{t_k}+y_{t_k}$ and revalue only the cash flows after $t_k$, conditional on $\{\tau>t_k\}$. The computations use a stylised quarterly grid, $t_k=k/4$ with maturity $T=17/4$, which approximates the contract's coupon dates (8 January, April, July and October) and its maturity (8 January 2019).
\begin{remark}[Fixed coupon and par]\label{rem:par}
Real CAT bonds pay the return on the collateral (a money-market rate) plus the spread, and it is for this floating structure that ``price at par when the coupon spread equals the market spread'' is exact. Our model has a fixed coupon. In the base case we approximate the floating coupon by $s_{t_k}+y_{t_k}$, where $y_{t_k}$ is the risk-free par yield for the remaining term, computed from the same Treasury curve, i.e.\ $y_{t_k}\Delta\sum_{j=k+1}^{m}P^{\mathrm{mkt}}(t_k,t_j)+P^{\mathrm{mkt}}(t_k,T)=1$. The approximation is exact in the absence of trigger risk: a floating coupon $\rho_j$ set at $t_{j-1}$ satisfies $\mathbb E^{\mathbb Q}[D(t_k,t_j)\rho_j\Delta\mid\mathcal F_{t_k}]=P^{\mathrm{mkt}}(t_k,t_{j-1})-P^{\mathrm{mkt}}(t_k,t_j)$, so both a floating-rate note and a fixed-rate bond paying $y_{t_k}$ are worth par, and the quoted spread $s_{t_k}$ is compensation for catastrophe risk only. With trigger risk the two coupon legs differ only through the weighting of the forward rates by the survival probabilities, a second-order effect when these probabilities are close to one. With the coupon equal to the spread (the convention of Li et al.\ \citeyearpar{Ref989800}), part of the spread is instead treated as compensation for the risk-free rate. This convention is reported as a robustness check (Section~\ref{sec:robust}); it understates the catastrophe premium whenever the term structure is not close to zero, and by 2018 it can no longer reproduce par.
\end{remark}

\noindent Parameters of the hypothetical CAT bond are summarised in Table~\ref{ref22}. Prices \eqref{ref18} are computed with the recursion \eqref{eq:surv}, which is exact up to discretisation, or by Monte Carlo simulation of the aggregate loss process.
\begin{table}[!h]
\centering
\caption{Selected parameters of the hypothetical CAT bond}
\label{ref22}
\begin{tabular}{l l r}
\hline
CAT bond parameters&\multicolumn{2}{r}{Values}\\
\hline
\multicolumn{2}{l}{$N$\quad Nominal amount}& $100$ \\
\multicolumn{2}{l}{$\zeta$ \quad Recovery fraction if triggered} & $0.5$ \\
\multicolumn{2}{l}{$T$ \quad Term (years)} & $17/4$ \\
\multicolumn{2}{l}{$D$ \quad Threshold level (aggregate, USD)} & $\XD\times10^{8}$ \\
\multicolumn{2}{l}{$\Delta$ \quad Time between coupon payments} & $0.25$ \\
\multicolumn{2}{l}{Coupon at valuation date $t_k$} & $s_{t_k}+y_{t_k}$ (Remark~\ref{rem:par}) \\
\hline
\end{tabular}
\end{table}

\subsection{Pooled analysis: na\"ive initialization}
We summarise results with two quantities. The pricing accuracy is measured by the root-mean-square pricing error over the quotation dates,
\[
\mathrm{RMSE}=\Big(\frac{1}{K+1}\sum_{k=0}^{K}\big(V^{\mathrm{model}}_{t_k}(\theta)-V^{\mathrm{mkt}}_{t_k}\big)^2\Big)^{1/2},
\]
in price units per 100 of nominal. The size of the catastrophe premium is measured by the implied \emph{risk multiple} $\ell(\theta)/\ell^{\mathbb P}$, the ratio of the risk-neutral trigger intensity of Proposition~\ref{prop:ident} to its physical counterpart $\ell^{\mathbb P}=\lambda\bar F(D)$: the market prices the bond as if a trigger event were $\ell(\theta)/\ell^{\mathbb P}$ times more likely than under the physical model. Unlike the split between $\lambda^\prime$ and $\kappa$, this quantity is identified by the prices (Proposition~\ref{prop:ident}), and it approximately equals the market's ratio of spread to expected loss (Remark~\ref{rem:multiple}).

We determine a first initial guess $\theta^{(0)}= ({\lambda^\prime}^{(0)}, \kappa^{(0)})$ by minimizing the sum of squared pricing errors,
\begin{eqnarray}\label{eq122}
\sum_{k=0}^{K} \left( V^{\text{model}}_{t_k}(s_{t_k}+y_{t_k};\theta) - 100 \right)^2 .
\end{eqnarray}
By Proposition~\ref{prop:ident}, \eqref{eq122} is nearly flat along the curve $\{\ell(\theta)=\ell^*\}$ of observationally equivalent measures. We minimise \eqref{eq122} with L-BFGS-B from $n=\XnRuns$ random starting points in the box $\Theta$ ($|\eta_f|,|\kappa|\le5$) and report, across runs, the fitted risk multiple, the parameters, the pricing errors and the entropy $\mathcal K_C(\theta^{(0)})$ (Table~\ref{tab:agg-summaries}). $\XnNonCorner$ runs converge to essentially the same interior point, $\lambda^\prime\approx\XlpLS$ and $\kappa\approx\XkaLS$ (risk multiple $\XmultLS$, RMSE $\XrmseMin$; the range of $\lambda^\prime$ across these runs is $\XlpSpreadLS$). The remaining $\XnCorner$ runs end at the corner $\lambda^\prime=e^{-5}-1$, $\kappa=-5$, where the trigger probability is negligible and prices hardly depend on $\theta$. The least-squares minimum thus combines a very large frequency loading with a negative severity loading, far from the physical measure ($\mathcal K_C\approx5.55$, against $\XKStar$ at the pooled optimum). It is determined by a second-order effect: with a large $\hat\lambda T$ ($\XlamHatTls$ at this point), the accumulation of many sub-threshold losses slightly changes the time profile of the survival probabilities, which the fit exploits. The na\"ive estimate is therefore driven by a feature of the loss model rather than by the identified trigger intensity; the entropy term of \eqref{eq:J} removes this dependence.

\begin{table}[htbp]
\centering\small
\caption{Na\"ive initialization: summaries across $\XnRuns$ runs (mean, standard deviation (SD), 25th percentile (P25), median and 75th percentile (P75)).}
\label{tab:agg-summaries}
\begin{tabular}{lccccc}
\toprule
 & Mean & SD & P25 & Median & P75 \\
\midrule
Pricing error (RMSE) & 1.562 & 0.826 & 1.418 & 1.418 & 1.418\\
$\ell(\theta^{(0)})/(\lambda\bar F(D))$ & 8.438 & 1.491 & 8.699 & 8.699 & 8.699\\
$\lambda^\prime$ & 24.01 & 4.42 & 24.78 & 24.78 & 24.78\\
$\kappa$ & -0.826 & 0.738 & -0.697 & -0.697 & -0.697\\
$\mathcal K_C(\theta^{(0)})$ & 5.38 & 0.94 & 5.55 & 5.55 & 5.55\\
\bottomrule

\end{tabular}
\end{table}

\subsection{Pooled analysis: optimal pricing measure}
We use the na\"ive estimate with the smallest pricing error as the initial value of Algorithm~\ref{alg:1}, which solves the pooled market-consistent entropy problem \eqref{eq:J}. We report: the optimal $\theta^*$; the implied risk multiple $\ell(\theta^*)/(\lambda\bar F(D))$; the entropy $\mathcal K_C(\theta^*)$; the in-sample pricing error; the number of iterations; the Q-price at the spread of 30 September 2014; the physical price \eqref{ref10912}; and the risk premium \eqref{eq:premium}. In addition, we check $\theta^*$ against a grid search over the two-dimensional parameter space.

\begin{table}[htbp]
\centering
\small
\caption{Pooled analysis: optimal parameters and CAT bond values at 30 September 2014: Q-price, physical price (CIR discounting, $\mathbb P$) and actuarial value $V^{\mathrm{act}}_{t_0}$ (market discounting, physical survival). Iterations: number of iterations of Algorithm~\ref{alg:1} until the stopping criterion is met.}
\label{tab:opt-params}
\begin{tabular}{lccccccc}
\toprule
\textbf{Initialization} & $\boldsymbol{\lambda'}$ & $\boldsymbol{\kappa}$ & $\mathcal K_C$ & \textbf{Iterations} & \textbf{Q-price} & \textbf{Physical} & \textbf{Actuarial} \\
\midrule
Na\"ive LS & 2.549 & 1.037 & 0.3275 & 27 & 99.694 & 113.666 & 107.876\\
\bottomrule

\end{tabular}
\end{table}

\begin{figure}[htbp]
\centering
\includegraphics[width=0.8\linewidth]{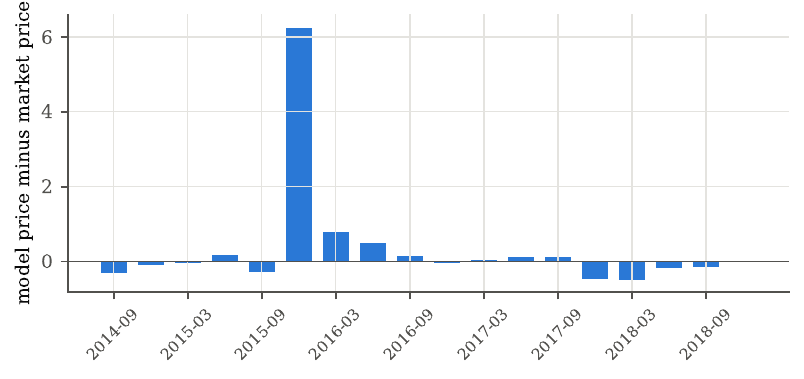}
\caption{Pooled analysis: model price at $\theta^*$ minus market price, per 100 of face value.}\label{fig:residuals}
\end{figure}

\noindent The observation-noise level of $0.25$ is not attainable: even the best fit has RMSE $\XrmseMin$, because a single time-homogeneous $\theta$ cannot follow the quarter-to-quarter variation of the spreads. The discrepancy principle with the estimated noise level then selects $\alpha=\Xalpha$, at which the sum of squared errors equals its target $n\hat\sigma^2=\XsseTarget$. Started from the na\"ive estimate, Algorithm~\ref{alg:1} converges in $\XitStar$ iterations (stopping when the norm of the Riemannian gradient falls below $\epsilon=10^{-5}$) to $\theta^*=(\lambda^{\prime*},\kappa^*)=(\XlpStar,\XkaStar)$, i.e.\ $\eta_f^*=\XetaStar$ (Table~\ref{tab:opt-params}). The solution agrees with an independent quasi-Newton solution to within \XdevRef, and with the polished maximiser of a $81\times81$ grid search over $(\eta_f,\kappa)$; together with the uniqueness check, this confirms that the maximiser is unique, so that other initialisations would lead to the same $\theta^*$. The entropy term moves the solution far from the na\"ive estimate: the risk multiple is $\XmultStar$, the entropy is $\mathcal K_C(\theta^*)=\XKStar$, and the in-sample RMSE is $\XrmseStar$. For comparison, the rule SSE $\le1.01\,\mathrm{SSE}_{\min}$ gives $\lambda^{\prime}=\XrelLp$ and $\kappa=\XrelKa$ (Table~\ref{tab:alpha}). At the spread of 30 September 2014 the Q-price is $\XqPrice$, the physical price \eqref{ref10912} is $\XpPrice$ and the actuarial benchmark is $\XactPrice$, so the risk premium is $\Pi_{t_0}=\XpremZero$. The physical price exceeds the actuarial value because it discounts with the CIR model under $\mathbb P$, whose expected short rates lie below the rates implied by the Treasury curve; this difference reflects the term structure, not catastrophe risk, which is why $\Pi_{t_k}$ is measured against the actuarial value. The pooled residuals (Figure~\ref{fig:residuals}) are dominated by 31 December 2015 ($\XresidDec$ per 100); all other residuals are at most $\XresidMaxOther$ in absolute value.

\begin{table}[htbp]
\centering\small
\caption{Pooled analysis: sensitivity of $\theta^*$ to the regularisation $\alpha$ ($\alpha_{\mathrm{base}}=\Xalpha$ from the discrepancy principle) and to the rule SSE $\le1.01\,$SSE$_{\min}$.}\label{tab:alpha}
\begin{tabular}{lccccccc}
\toprule
 & $\alpha$ & $\lambda^{\prime*}$ & $\kappa^*$ & $\eta_f^*$ & multiple & $\mathcal K_C(\theta^*)$ & RMSE\\
\midrule
0.1\,$\alpha_{\mathrm{base}}$ & 8.321 & 2.804 & 1.054 & 1.336 & 9.96 & 0.3733 & 1.426\\
0.3\,$\alpha_{\mathrm{base}}$ & 24.96 & 2.736 & 1.055 & 1.318 & 9.79 & 0.3626 & 1.433\\
1\,$\alpha_{\mathrm{base}}$ (base) & 83.21 & 2.549 & 1.037 & 1.267 & 9.20 & 0.3275 & 1.504\\
3\,$\alpha_{\mathrm{base}}$ & 249.6 & 2.099 & 0.980 & 1.131 & 7.76 & 0.2460 & 1.932\\
10\,$\alpha_{\mathrm{base}}$ & 832.1 & 1.140 & 0.776 & 0.761 & 4.66 & 0.0962 & 3.448\\
SSE$\le1.01\,$SSE$_{\min}$ & 0.9212 & 3.005 & 0.980 & 1.388 & 10.03 & 0.3803 & 1.425\\
\bottomrule

\end{tabular}
\end{table}

\subsection{Date-wise analysis}\label{sec:datewise}
For each of the 18 quarter-ends we solve the date-wise problem \eqref{eq:DW} with Algorithm~\ref{alg:2}, using only the spread and the Treasury curve observed at that date. The outputs are, for each $t_k$:
the implied risk-neutral trigger intensity $\ell^*_k=\ell(\theta^*_k)$ and the risk multiple $\ell^*_k/(\lambda\bar F(D))$;
the entropy-selected premia $\lambda^{\prime*}_k$ and $\kappa^*_k$;
the minimal relative entropy $\mathcal K_C(\theta^*_k)$, a measure of how far the market's pricing of catastrophe risk is from the physical model at that date;
and the risk premium $\Pi_{t_k}$ of \eqref{eq:premium}.
We validate Algorithm~\ref{alg:2} against a one-dimensional search along the feasible curve and against the closed-form condition \eqref{eq:foc}. We also compare the date-wise series with the pooled estimate: a pooled $\theta^*$ that lies well inside the range of $\{\theta^*_k\}$, with small and unpatterned pooled residuals, supports a constant premium; systematic residuals point to time-varying premia. Two dates require care: the year-end spike of 31 December 2015 (Section~\ref{ref54}) and the quote of 31 December 2018, eight days before maturity. We report results with and without them.


\begin{table}[htbp]
\centering\small
\caption{Date-wise analysis.}
\label{tab:datewise}
\begin{tabular}{lccccccc}
\toprule
$t_k$ & $s_{t_k}$ (\%) & $\ell^*_k/(\lambda\bar F(D))$ & $\lambda^{\prime*}_k$ & $\kappa^*_k$ & $\mathcal K_C(\theta^*_k)$ & $\Pi_{t_k}$ & freq.\ share (\%)\\
\midrule
2014-09-30 & 2.16 & 8.867 & 2.437 & 1.029 & 0.3081 & 7.876 & 56.6\\
2014-12-31 & 2.23 & 9.102 & 2.511 & 1.037 & 0.3029 & 7.699 & 56.9\\
2015-03-31 & 2.26 & 9.189 & 2.540 & 1.039 & 0.2885 & 7.370 & 57.0\\
2015-06-30 & 2.32 & 9.433 & 2.618 & 1.047 & 0.2811 & 7.088 & 57.3\\
2015-09-30 & 2.18 & 8.827 & 2.431 & 1.024 & 0.2339 & 6.155 & 56.6\\
2015-12-31 & 4.56 & 18.260 & 5.366 & 1.224 & 0.6712 & 12.620 & 63.7\\
2016-03-31 & 2.60 & 10.449 & 2.940 & 1.076 & 0.2609 & 6.354 & 58.4\\
2016-06-30 & 2.50 & 10.046 & 2.816 & 1.063 & 0.2226 & 5.551 & 58.0\\
2016-09-30 & 2.36 & 9.477 & 2.640 & 1.044 & 0.1821 & 4.683 & 57.5\\
2016-12-31 & 2.27 & 9.110 & 2.527 & 1.031 & 0.1517 & 3.977 & 57.1\\
2017-03-31 & 2.33 & 9.329 & 2.597 & 1.037 & 0.1380 & 3.585 & 57.3\\
2017-06-30 & 2.38 & 9.509 & 2.655 & 1.043 & 0.1221 & 3.149 & 57.6\\
2017-09-30 & 2.41 & 9.621 & 2.692 & 1.046 & 0.1037 & 2.666 & 57.7\\
2017-12-31 & 1.83 & 7.304 & 1.968 & 0.952 & 0.0521 & 1.560 & 54.7\\
2018-03-31 & 1.65 & 6.576 & 1.741 & 0.913 & 0.0325 & 1.037 & 53.5\\
2018-06-30 & 1.96 & 7.795 & 2.124 & 0.974 & 0.0292 & 0.847 & 55.5\\
2018-09-30 & 1.69 & 6.719 & 1.788 & 0.920 & 0.0113 & 0.357 & 53.8\\
2018-12-31$^\ddagger$ & 3.11 & -- & -- & -- & -- & 0.000 & --\\
\bottomrule

\end{tabular}
\par\smallskip\footnotesize Coupon $s_{t_k}+y_{t_k}$, aggregate trigger, $D=\XD\times10^8$, $\zeta=0.5$. The frequency share is $100\,\eta_f^*/\ln(\ell^*_k/\lambda\bar F(D))$. $^\ddagger$Maturity date on the stylised grid; the price equals 100 for every $\theta$.
\end{table}

\begin{figure}[htbp]
\centering
\includegraphics[width=0.9\linewidth]{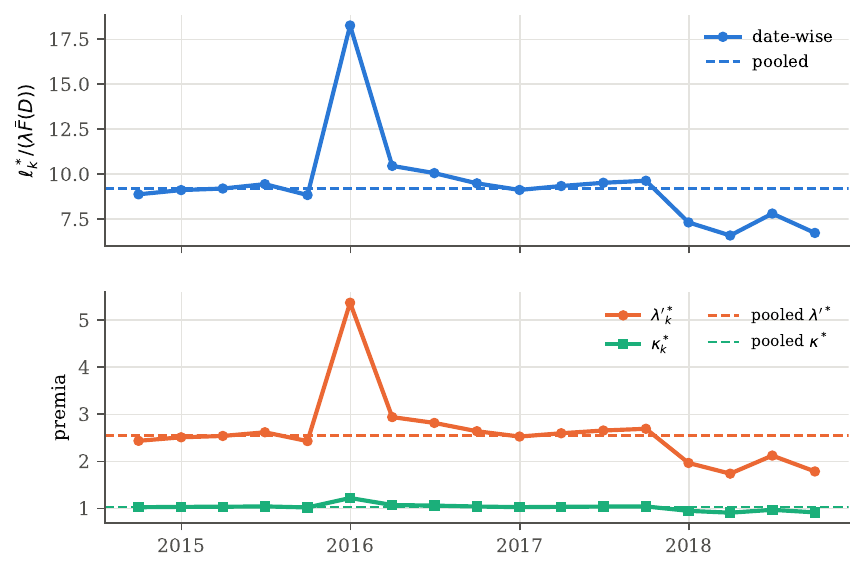}
\caption{Date-wise implied risk multiple (top) and entropy-selected premia $\lambda^{\prime*}_k$ and $\kappa^*_k$ (bottom) at the feasible quote dates; dashed lines are the pooled estimates.}
\label{fig:datewise}
\end{figure}

\noindent Table~\ref{tab:datewise} and Figure~\ref{fig:datewise} report the date-wise results. Algorithm~\ref{alg:2} converged at every date before maturity in $\XitDWmin$--$\XitDWmax$ iterations, and its output agrees with the one-dimensional search along the feasible curve to within \Xdevone. The closed-form condition \eqref{eq:foc}, which is exact only for a per-occurrence trigger, reproduces the split to within $\XdevOcc$. The date-wise problem is feasible at all $\XnFeas$ quote dates before maturity; with the coupon equal to the spread it would be feasible at only $\XliFeas$ of them (Remark~\ref{rem:par}). The maturity date $t_{17}$ is uninformative.

\begin{remark}[Risk multiple and spread-to-expected-loss ratio]\label{rem:multiple}
To first order, the implied risk multiple equals the market's spread-to-expected-loss ratio. With the coupon $s_{t_k}+y_{t_k}$ and survival probabilities close to $e^{-\ell t}$, the par condition reads, to first order in the spreads and the risk-free rates, $\sum_{j=k+1}^{m}P^{\mathrm{mkt}}(t_k,t_j)\Delta\,\{s_{t_k}-\ell^*_k(1-\zeta)\}\approx0$, so that $\ell^*_k(1-\zeta)\approx s_{t_k}$. The physical annual expected principal loss is $\mathrm{EL}=\lambda\bar F(D)(1-\zeta)=\XEL\%$, which matches the reference bond's expected loss of 0.25\%, so that $\ell^*_k/(\lambda\bar F(D))\approx s_{t_k}/\mathrm{EL}$. At issuance the date-wise multiple is $\XmultZero$ against $s_{t_0}/\mathrm{EL}=\XspreadEL$, and the same agreement holds at the other dates (e.g.\ $4.56/\XEL\approx18.6$ against $\XmultDec$ at the end of 2015, and $1.65/\XEL\approx6.7$ against $\XmultMinEx$ in March 2018). The model therefore reproduces the multiple in which the market itself quotes these bonds. What it adds is a decomposition of this multiple into frequency and severity loadings; by Proposition~\ref{prop:ident}, this decomposition is not determined by the quotes but by the entropy criterion, i.e.\ it is the split closest to the physical measure among those consistent with the prices.
\end{remark}

\begin{figure}[htbp]
\centering
\includegraphics[width=0.68\linewidth]{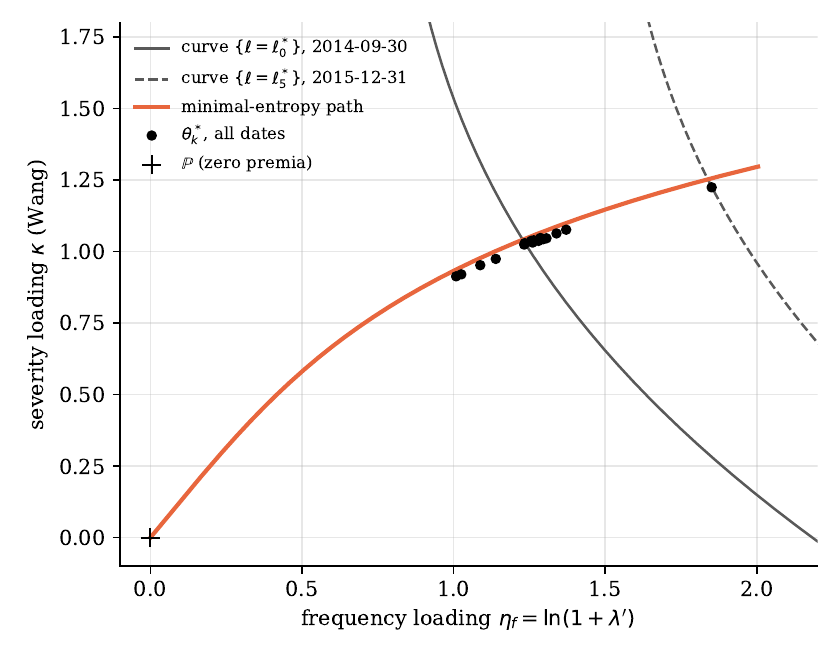}
\caption{Date-wise entropy-selected measures in the coordinates $(\eta_f,\kappa)$, the minimal-entropy path of Corollary~\ref{cor:path} (per-occurrence approximation) and the curves of observationally equivalent measures for two dates.}\label{fig:path}
\end{figure}

\subsection{Robustness and sensitivity}\label{sec:robust}
All pooled and date-wise results are recomputed under the following variations:
(i) coupon convention, the base $s_{t_k}+y_{t_k}$ versus the coupon equal to the spread, $s_{t_k}$ (Li et al.\ \citeyearpar{Ref989800}; Remark~\ref{rem:par});
(ii) discounting, market discount factors versus the CIR model with $\lambda_r$ calibrated to the curve;
(iii) threshold, the matched $D=\XD\times10^8$ versus $D=6\times10^7$;
(iv) recovery fraction $\zeta\in\{0.25,0.5,0.75\}$;
(v) severity shape $a$ and intensity $\lambda$, each varied by $\pm20\%$ with $D$ held fixed (Safarveisi et al.\ \citeyearpar{Ref8889} report point estimates only, so confidence intervals are not available);
(vi) the regularisation $\alpha$ in \eqref{eq:J} (Table~\ref{tab:alpha});
(vii) exclusion of the 31 December 2015 and 31 December 2018 quotes.

\begin{table}[htbp]
\centering\scriptsize
\setlength{\tabcolsep}{3pt}
\caption{Robustness: pooled optimum and date-wise summaries under the variations (i)--(v) and (vii); the variations of $\alpha$ are in Table~\ref{tab:alpha}.}\label{tab:robust}
\begin{tabular}{lcccccccc}
\toprule
 & \multicolumn{5}{c}{Pooled} & \multicolumn{3}{c}{Date-wise}\\
\cmidrule(lr){2-6}\cmidrule(lr){7-9}
Variation & $\lambda^{\prime*}$ & $\kappa^*$ & multiple & RMSE & Q-price & feasible & mean multiple & freq.\ share (\%)\\
\midrule
Base (coupon $C+y$) & 2.549 & 1.037 & 9.20 & 1.504 & 99.69 & 17 & 9.39 & 57.0\\
(i) coupon = spread (Li et al.) & 1.030 & 0.738 & 4.30 & 1.736 & 98.12 & 14 & 5.05 & 49.2\\
(ii) CIR discounting & 2.534 & 1.035 & 9.15 & 1.530 & 99.60 & 17 & 9.36 & 57.0\\
(iii) $D=6\times10^7$ & 2.228 & 0.674 & 5.13 & 1.503 & 99.69 & 17 & 5.24 & 71.6\\
(iv) $\zeta$ = 0.25 & 1.593 & 0.890 & 6.13 & 1.545 & 99.66 & 17 & 6.27 & 52.6\\
(iv) $\zeta$ = 0.75 & 5.424 & 1.227 & 18.45 & 1.388 & 99.79 & 17 & 18.64 & 63.7\\
(v) $a$ -20\% & 2.629 & 1.168 & 11.57 & 1.504 & 99.70 & 17 & 11.80 & 52.6\\
(v) $a$ +20\% & 2.475 & 0.939 & 7.79 & 1.504 & 99.69 & 17 & 7.95 & 60.6\\
(v) $\lambda$ -20\% & 3.260 & 1.109 & 11.51 & 1.504 & 99.69 & 17 & 11.75 & 59.2\\
(v) $\lambda$ +20\% & 2.073 & 0.972 & 7.66 & 1.504 & 99.70 & 17 & 7.81 & 55.1\\
(vii) excl. 2015-12-31, 2018-12-31 & 2.512 & 1.032 & 9.08 & 0.335 & 99.80 & 16 & 8.83 & 56.6\\
\bottomrule

\end{tabular}
\par\smallskip\footnotesize Q-price at the spread of 30 September 2014. Date-wise means over the feasible dates before maturity; the frequency share is $100\,\eta_f^*/\ln(\ell^*/\lambda\bar F(D))$.
\end{table}

\noindent\textbf{Discussion of results.} (i) \emph{Initialisation.} The na\"ive least-squares estimate lies far from the physical measure and is driven by a second-order feature of the loss model, but Algorithm~\ref{alg:1} started from it reaches the same pooled $\theta^*$ as an independent solver and the grid search; because the maximiser of \eqref{eq:J} is unique, the choice of initialisation affects only the computing time. (ii) \emph{Sign and size of the premium.} The risk premium $\Pi_{t_k}$ is positive at every quote date, between $\XpiMin$ and $\XpiMax$ per 100, and the implied trigger intensity exceeds its physical value throughout, in line with the evidence that CAT bond spreads exceed expected losses (Braun \citeyear{Ref9000}). The risk multiple matches the spread-to-expected-loss ratio (Remark~\ref{rem:multiple}). (iii) \emph{Behaviour over time.} The date-wise risk multiple lies between $\XmultMinEx$ and $\XmultMaxEx$ at all dates other than 31 December 2015, where it peaks at $\XmultDec$; it is close to the pooled value $\XmultStar$ until 2017 and lower in 2018, when the quoted spreads fell. A single time-homogeneous measure fits the quotes well apart from the spike of December 2015: excluding that date and the last quote lowers the RMSE from $\XrmseStar$ to $\XexRmse$ per 100. (iv) \emph{The premium split.} Both premia are loaded at every date: $\lambda^{\prime*}_k$ ranges from $\XlpMinDW$ to $\XlpMaxDW$ and $\kappa^*_k$ from $\XkaMinDW$ to $\XkaMaxDW$, and the frequency loading accounts for $\XfsMin$--$\XfsMax\%$ of the log risk multiple. By Corollary~\ref{cor:path}, the date-wise solutions lie on the minimal-entropy path (Figure~\ref{fig:path}), so the co-movement of the two premia is implied by the single identified quantity $\ell^*_k$ and carries no further information. The split also moves with the trigger level: with $D=6\times10^7$ the frequency share rises to about 72\% (Table~\ref{tab:robust}). It should therefore be read as the allocation implied by the minimal-entropy principle for the given trigger level, not as a feature revealed by the market data. (v) \emph{Regularisation.} Across a hundredfold range of $\alpha$ the pooled multiple lies between $\XalphaMultMin$ and $\XalphaMultMax$ and both premia remain positive (Table~\ref{tab:alpha}). For $\alpha$ up to its base value the multiple stays between $\XmultStar$ and about 10; larger values shrink the measure towards $\mathbb P$ at the cost of a much poorer fit. (vi) \emph{Robustness.} The level of the implied risk multiple depends mainly on the assumptions that fix the cash flows and the loss given trigger (Table~\ref{tab:robust}). Against $\XmultStar$ in the base case, it is $\XliMult$ with the coupon equal to the spread, which treats part of the spread as compensation for the risk-free rate; $\XcirMult$ with CIR discounting; between $\XzLoMult$ ($\zeta=0.25$) and $\XzHiMult$ ($\zeta=0.75$) for the recovery fraction; and $\XdAltMult$ with $D=6\times10^7$, for which the physical trigger probability is higher. The $\pm20\%$ variations of $a$ and $\lambda$ give $\XvMin$--$\XvMax$. In every variation both premia are loaded, and the average date-wise frequency share stays between $\XfsRobMin\%$ and $\XfsRobMax\%$.

\section{Conclusion}
In this paper, we propose an end-to-end framework for pricing catastrophe (CAT) bonds in incomplete markets. The procedure begins with the specification of a pricing model and culminates in the selection of an optimal pricing measure based on the principle of financial entropy, combined with the information contained in observed secondary-market spreads, both pooled over the life of the bond and date by date. The model incorporates a Cox--Ingersoll--Ross (CIR) process for interest-rate dynamics and a compound Poisson process with generalized Pareto severities for catastrophe losses, with a Wang distortion for the severity premium. We show that the entropy principle alone would select the physical measure, that the interest-rate premium is fixed by the term structure, and that the frequency and severity premia are only weakly identified by CAT bond prices. The entropy criterion resolves exactly this weakly identified direction, the selected measures lie on a single minimal-entropy path determined by the identified trigger intensity, and the financial entropy and Riemannian metric are available in closed form. The value under the physical measure is obtained through a Feynman--Kac argument.

In the application to Golden State Re II, market prices imply a positive catastrophe premium at every quarter-end from 2014 to 2018. The entropy-selected pooled measure implies a risk-neutral trigger intensity $\XmultStar$ times its physical value, with a frequency loading $\lambda^{\prime*}=\XlpStar$ and a severity loading $\kappa^*=\XkaStar$; the date-wise risk multiple lies between $\XmultMinEx$ and $\XmultMaxEx$, with a peak of $\XmultDec$ at the end of 2015. This multiple is the model-consistent counterpart of the market's spread-to-expected-loss ratio. The split between frequency and severity loadings depends on the trigger level and follows the identified intensity, so it should be read as the allocation implied by the selection principle rather than as a market-revealed quantity. The translation of quoted spreads into prices also matters: setting the coupon equal to the spread, as is sometimes done, treats part of the spread as compensation for the risk-free rate and understates the risk multiple by about one half ($\XliMult$ against $\XmultStar$).

Two implications follow. First, calibrations that fit several distinct catastrophe premia to the price history of a single bond should report which combinations of these premia are actually identified. Second, statements about the frequency/severity composition of catastrophe risk premia require additional market information; bonds on the same peril with different attachment points are a natural source, because they observe the risk-neutral loss distribution at different levels, and exploiting this cross-section is left for future research. Further limitations include the parametric family of measures, the basis between the physical loss model and the sponsor's modelled loss index, the stylised coupon grid, the approximation of the floating coupon, and the use of exponential utility; extensions to other utility functions, time-varying premia and dependence between insurance and financial risks are left for future work.

\bibliographystyle{agsm}

\appendix
\renewcommand{\theequation}{\thesection.\arabic{equation}}
\section{Zero-coupon bond price}
\noindent The risk-free savings account is described by the equation $dB(t)= r(t)B(t)dt$, that is,\ $B(t) = B(0)\exp\{\int_{0}^{t}r(u) du\}$ with $r$ given by (\ref{ref3}). The discount factor between $t$ and $T$ is
\begin{eqnarray}\label{ref72}
D(t, T) = \frac{B(t)}{B(T)} = \exp\bigg\{-\int_{t}^{T}r(u)du\bigg\}.
\end{eqnarray}
Let $P(t, T)$ denote the time-$t$ price of a zero-coupon bond with nominal value 1 at maturity $T$, so $P(T, T)= 1$. By the risk-neutral pricing principle, $\{P(t, T)/B(t)\}$ is a $\mathbb{Q}$-martingale, hence
\begin{eqnarray}\label{ref4}
P(t, T) = \mathbb E^{\mathbb{Q}}\bigg[\frac{B(t)}{B(T)}\,\Big|\, \mathcal{F}_t\bigg] = \mathbb E^{\mathbb{Q}}\big[D(t, T)\,|\, \mathcal{F}_t\big].
\end{eqnarray}
Under the $\mathbb Q$-dynamics \eqref{eq:cirQ} the analytical solution is (see, e.g., Brigo and Mercurio \citeyear{Ref6})
\begin{eqnarray}\label{ref101}
P(t, T, r(t), \theta^{*}_{r} , m^{*}_{r},\sigma_r) = A(t , T)e^{-B(t , T)r(t)},
\end{eqnarray}
\begin{eqnarray*}
A(t , T)  = \left[\frac{2\gamma e^{(\gamma+\theta^{*}_{r})(T-t)/2}}{(\gamma + \theta^{*}_{r})(e^{\gamma(T-t)}-1)+2\gamma}\right]^{\frac{2m^{*}_{r}\theta^{*}_{r}}{\sigma_r^2}}, \\
B(t, T) = \frac{2(e^{\gamma(T-t)}-1)}{(\gamma + \theta^{*}_{r})(e^{\gamma(T-t)}-1)+2\gamma}, \\
\gamma = \sqrt{(\theta^{*}_{r})^2 + 2\sigma^2_r}.
\end{eqnarray*}
(Here $\gamma$ is the usual CIR constant and should not be confused with the Girsanov kernel of Section~\ref{section2.3}.)

\section{The Feynman--Kac formula and proof of Lemma \ref{ref85}}
\begin{theorem}[Feynman--Kac formula]\label{ref80}
Let $\{X_t\}$ be an It\^o diffusion with drift $b(x)$ and diffusion coefficient $\sigma(x)$, let $f$ and $q\ge0$ be continuous, and let
\begin{equation*}
v(t, x) = \mathbb{E}\Big[f(X_t)\exp\Big(-\int_{0}^{t}q(X_s)ds\Big)\Big| X_0 = x\Big].
\end{equation*}
Under suitable regularity conditions (e.g.\ Gan et al.\ \citeyear{Ref24}; {\O}ksendal \citeyear{Ref35}), $v$ solves the Cauchy problem
\begin{eqnarray}\label{ref78}
\frac{\partial v}{\partial t}(t, x) = \mathcal{L}v(t, x) - q(x)v(t, x),\quad t>0, \qquad v(0, x) = f(x),
\end{eqnarray}
with $\mathcal{L}v = b(x)\,\partial_x v +\frac{1}{2}\sigma^2(x)\,\partial_{xx} v$. Conversely, a sufficiently regular and suitably bounded solution of \eqref{ref78} coincides with $v$.
\end{theorem}
\noindent We now consider the CIR process as defined in (\ref{ref3}), and find its corresponding solution of the partial differential equation (\ref{ref78}).
\begin{lemma}\label{ref81}
Let $g(t, x)$ be a function defined on $[0, T]\times (0, +\infty)$ that satisfies the CIR partial differential equation (PDE) with $b(x) = \bar{\theta}_r(\bar{m}_r - x)$ and $\sigma(x) = \sigma_r \sqrt{x}$. In other words, it is a solution of the following deterministic PDE
\begin{eqnarray}\label{ref79}
\begin{gathered}[b]
\frac{\partial g}{\partial t}(t, x) = \frac{1}{2}\sigma^2_r x \frac{\partial^2 g }{\partial x^2}(t, x) + \bar{\theta}_r (\bar{m}_r - x)\frac{\partial g}{\partial x}(t, x)-q(x)g(t, x),\\ (t, x) \in (0, T]\times \mathbb{R}^+,\\
g(0, x) = f(x), \quad x\in \mathbb{R}^+.
\end{gathered}
\end{eqnarray}
If we choose $f(x) = e^{-ux}$ and $q(x) = wx$ with $u,w\ge0$, then the time-$T$ value of the function $g(t, x)$ can be written as
\begin{eqnarray}\label{ref8199}
g(T, x) = \exp\big\{\mathcal A(T)-\mathcal B(T)\,x\big\},
\end{eqnarray}
where $\mathcal A(T)$ and $\mathcal B(T)$ are given in Lemma~\ref{ref85}.
\end{lemma}
\begin{proof}
We need to find the solution of the PDE given by (\ref{ref79}). To this end, we start with an initial guess. One reasonable choice would be $g(t, x) = K(t)e^{-\mathcal B(t)x}$, where $K(t)$ and $\mathcal B(t)$ are deterministic functions to be specified. As we know that $g(0, x) = e^{-ux}$, we have that $K(0) = 1$ and $\mathcal B(0)=u$. Now, if $g(t, x) = K(t)e^{-\mathcal B(t)x}$ is a solution of (\ref{ref79}), then $K(t)$ and $\mathcal B(t)$ have to satisfy
\begin{eqnarray}
\begin{gathered}
K^{\prime}(t)e^{-\mathcal B(t)x}-\mathcal B^{\prime}(t)\,x\,K(t)e^{-\mathcal B(t)x} = \frac{1}{2}\sigma^2_r x K(t)\mathcal B(t)^2e^{-\mathcal B(t)x}\\
-\bar{\theta}_r(\bar{m}_r - x)K(t)\mathcal B(t)e^{-\mathcal B(t)x} - wxK(t)e^{-\mathcal B(t)x},
\end{gathered}
\end{eqnarray}
or simply, after dividing by $K(t)e^{-\mathcal B(t)x}$,
\begin{eqnarray}
\frac{K^{\prime}(t)}{K(t)}-\mathcal B^{\prime}(t)\,x = \frac{\sigma^2_r \mathcal B(t)^2}{2}\,x-\bar{\theta}_r \bar{m}_r\mathcal B(t)+\bar\theta_r\mathcal B(t)\,x-wx ,
\end{eqnarray}
where $K^{\prime}(t) = \frac{dK(t)}{dt}$ and $\mathcal B^{\prime}(t) = \frac{d\mathcal B(t)}{dt}$. Since this identity must hold for every $x>0$, the terms that do not depend on $x$ and the terms that are proportional to $x$ must be equal separately. This gives two ordinary differential equations:
\begin{eqnarray}
\mathcal B^\prime(t)&=&w-\bar\theta_r\mathcal B(t)-\frac12\sigma_r^2\mathcal B(t)^2,\qquad \mathcal B(0)=u, \label{eq:ricB}\\
K^\prime(t)&=&-\bar\theta_r\bar m_r\,\mathcal B(t)\,K(t),\qquad K(0)=1. \label{eq:odeK}
\end{eqnarray}
Note that the coefficient of $x$ in the exponent has to be a general function $\mathcal B(t)$: the simpler choice $\mathcal B(t)=wt+u$, i.e.\ the guess $K(t)e^{-wxt-ux}$, satisfies \eqref{eq:ricB} only when $\bar\theta_r=\sigma_r=0$; otherwise the terms proportional to $x$ do not cancel and no function $K(t)$ of $t$ alone solves the resulting equation.

\noindent Using separation of variables technique for \eqref{eq:ricB}, with $\varrho=\sqrt{\bar\theta_r^2+2\sigma_r^2w}$ and $b_\pm=(-\bar\theta_r\pm\varrho)/\sigma_r^2$ the roots of $\frac12\sigma_r^2b^2+\bar\theta_rb-w=0$ (so that $w-\bar\theta_r\mathcal B-\frac12\sigma_r^2\mathcal B^2=-\frac12\sigma_r^2(\mathcal B-b_+)(\mathcal B-b_-)$), we have that
\begin{eqnarray*}
\mathcal B^{\prime}(t) &=& -\frac12\sigma_r^2\big(\mathcal B(t)-b_+\big)\big(\mathcal B(t)-b_-\big), \\
&\Rightarrow & \frac{d\mathcal B(t)}{\big(\mathcal B(t)-b_+\big)\big(\mathcal B(t)-b_-\big)} = -\frac12\sigma_r^2\,dt \\
&\Rightarrow & \int_{0}^{T}\Big[\frac{1}{\mathcal B(t)-b_+}-\frac{1}{\mathcal B(t)-b_-}\Big]d\mathcal B(t) = -\frac{\sigma_r^2}{2}(b_+-b_-)\,T \\
&\Rightarrow & \ln\frac{\mathcal B(T)-b_+}{\mathcal B(T)-b_-}-\ln\frac{u-b_+}{u-b_-} = -\varrho\,T\\
&\Rightarrow & \mathcal B(T)=\frac{b_+-c\,e^{-\varrho T}\,b_-}{1-c\,e^{-\varrho T}},\qquad c=\frac{u-b_+}{u-b_-}.
\end{eqnarray*}
Similarly, using separation of variables technique for \eqref{eq:odeK}, we have that
\begin{eqnarray*}
K^{\prime}(t) &=& -\bar\theta_r\bar m_r\,\mathcal B(t)\,K(t), \\
&\Rightarrow & \frac{dK(t)}{K(t)} = -\bar\theta_r\bar m_r\,\mathcal B(t)\,dt \\
&\Rightarrow & \int_{0}^{T}\frac{dK(t)}{K(t)} = -\bar\theta_r\bar m_r\int_{0}^{T}\mathcal B(t)\,dt \\
&\Rightarrow & \ln K(T) - \ln K(0) = -\bar\theta_r\bar m_r\Big[b_+T+\frac{2}{\sigma_r^2}\ln\frac{1-c\,e^{-\varrho T}}{1-c}\Big]\\
&\Rightarrow & K(T) = \exp\{\mathcal A(T)\},\qquad \mathcal A(T)=-\bar\theta_r\bar m_r\Big[b_+T+\frac{2}{\sigma_r^2}\ln\frac{1-c\,e^{-\varrho T}}{1-c}\Big],
\end{eqnarray*}
where the integral of $\mathcal B$ follows from $\mathcal B(t)=b_-+(b_+-b_-)/(1-c\,e^{-\varrho t})$, $\int_0^T\frac{dt}{1-c\,e^{-\varrho t}}=T+\frac1\varrho\ln\frac{1-c\,e^{-\varrho T}}{1-c}$ and $(b_+-b_-)/\varrho=2/\sigma_r^2$. So, the solution at time $t=T$ is
\begin{eqnarray*}
g(T, x) = K(T)e^{-\mathcal B(T)x} = \exp\big\{\mathcal A(T)-\mathcal B(T)\,x\big\}.
\end{eqnarray*}
For $u,w\ge0$ we have $b_-<0\le b_+$ and $|c|<1$, so all expressions are well defined; moreover $\mathcal B(t)\ge0$ and $\mathcal A(t)\le0$, so that $0<g(t,x)\le1$.
\end{proof}

\begin{proof}[Proof of Lemma \ref{ref85}]
We assume that $f(x) = e^{-ux}$ and $q(x) = wx$. Then by Theorem \ref{ref80}, the classical solution of the following Cauchy problem
\begin{eqnarray}\label{ref82}
\begin{gathered}[b]
\frac{\partial v}{\partial t}(t, r_0) = \frac{1}{2}\sigma^2_r r_0 \frac{\partial^2 v }{\partial r_0^2}(t, r_0) + \bar{\theta}_r (\bar{m}_r - r_0)\frac{\partial v}{\partial r_0}(t, r_0)-q(r_0)v(t, r_0),\\ (t, r_0) \in (0, T]\times \mathbb{R}^+,\\
v(0, r_0) = f(r_0), \quad r_0\in \mathbb{R}^+,
\end{gathered}
\end{eqnarray}
admits the following representation at time $t=T$
\begin{eqnarray}
v(T, r_0) = \mathbb{E}^{\mathbb{P}}\left[\exp\Big\{-w\int_{0}^{T} r(s)ds - u\,r(T)\Big\}\Big |r(0)= r_0\right].
\end{eqnarray}
But, by Lemma \ref{ref81}, the function $g(T,r_0)=\exp\{\mathcal A(T)-\mathcal B(T)\,r_0\}$ also satisfies the Cauchy problem (\ref{ref82}), and it is smooth and bounded ($0<g\le1$). Hence, Theorem \ref{ref80} gives $g(T,r_0 ) = v(T, r_0)$. Since the diffusion coefficient of the CIR process vanishes at zero, this identification can also be verified directly: for fixed $T$, $M_s=\exp\{-w\int_0^s r(v)\,dv\}\,g(T-s,r(s))$, $0\le s\le T$, is by It\^o's formula and (\ref{ref82}) a local martingale with $0<M_s\le1$, hence a martingale, so that $g(T,r_0)=M_0=\mathbb E^{\mathbb P}[M_T]=v(T,r_0)$.

\noindent Finally, for $u=0$ and $w=1$ one has $c=b_+/b_-$, and a direct simplification gives $\mathcal B(T)=B(0,T)$ and $e^{\mathcal A(T)}=A(0,T)$ in (\ref{ref101}) with $(\theta_r^*,m_r^*)$ replaced by $(\bar\theta_r,\bar m_r)$, i.e.\ the CIR zero-coupon bond price under $\mathbb P$. Numerically, for the parameters of Table~\ref{ref20} the two formulas agree to machine precision, and for $u>0$ the formula agrees with Monte Carlo simulation within sampling error.
\end{proof}
\noindent Corollary~\ref{ref88} follows from the Markov property of $r$ and the time-homogeneity of \eqref{ref3}.

\section{The geometric framework of Gao et al.\ (2004)}\label{app:gln}
For completeness, we recall the construction of Gao et al.\ \citeyearpar{Ref24001}. In their Brownian setting, the $n$-dimensional manifold $\mathcal{S}$ of problem \eqref{ref1002} is
\begin{eqnarray}
 \left\{ \mathbb{Q}_\theta \sim \mathbb{P} : \frac{d\mathbb{Q}_\theta}{d\mathbb{P}} =\mathcal{E}\left(\int_{0}^{T}\eta_s d\mathcal{W}(s)\right) = \mathcal{E} \left( \int_{0}^{T} \sum_{i=1}^{n} \theta_i \xi_s^i \, d\mathcal{W}(s) \right),\; \theta \in \mathbb{R}^n \right\}.
\end{eqnarray}
Here, $\theta = (\theta_1, \dots, \theta_n)$ is a coordinate vector, $\mathbb{Q}_{\theta}$ is the corresponding point on $\mathcal{S}$, $\mathcal{E}(\cdot)$ denotes the Dol\'eans-Dade exponential, and $\{\xi_s^i\}_{i=1}^n$ are known, nontrivial stochastic processes, often involving multiple iterated It\^o integrals, satisfying appropriate regularity conditions. The summation term represents the truncated Wiener--It\^o chaos expansion of the incomplete-market price of risk $\eta_s=\sum_{i=1}^{n} \theta_i \xi_s^i$ associated with the non-hedgeable risk factor $\mathcal{W}(s)$. For example, consider a two-dimensional Brownian motion $(B(t),\mathcal{W}(t))$, where $B$ drives the traded (hedgeable) security, with market price of risk $\phi_s$ fixed by no-arbitrage, and $\mathcal{W}$ is the non-hedgeable factor with market price of risk $\eta_s$. In this case,
\begin{eqnarray}
 \mathcal S&=&\left\{ \mathbb{Q}_\theta \sim \mathbb{P} : \frac{d\mathbb{Q}_\theta}{d\mathbb{P}}  = \exp\Big\{\int_{0}^{T}\phi_s dB(s) - \frac{1}{2}\int_{0}^{T}\phi_s^2 ds \right.\nonumber\\
&&\left.\qquad + \int_{0}^{T}\sum_{i=1}^{n} \theta_i \xi_s^i d\mathcal{W}(s)-\frac{1}{2}\int_{0}^{T}\Big(\sum_{i=1}^{n} \theta_i \xi_s^i\Big)^2 ds\Big\},\; \theta \in \mathbb{R}^n \right\}. \nonumber
\end{eqnarray}

\section{Proofs of Section \ref{refsection3}}\label{app:proofs}
\paragraph{Proposition \ref{prop:KL}.} By the independence of $W$ and $L$ under $\mathbb Q$, $\mathrm{KL}(\mathbb Q\|\mathbb P)=\mathbb E^{\mathbb Q}[\ln Z_{1,T}]+\mathbb E^{\mathbb Q}[\ln Z_{2,T}]$. For the jump part, $\mathbb E^{\mathbb Q}[\ln Z_{2,T}]=-\lambda\lambda^\prime T+\mathbb E^{\mathbb Q}[N_T]\ln(1+\lambda^\prime)+\mathbb E^{\mathbb Q}[N_T]\,\mathbb E^{\mathbb Q}[\kappa Z_1-\kappa^2/2]$, where $\mathbb E^{\mathbb Q}[N_T]=\lambda(1+\lambda^\prime)T$ and $Z_1\sim N(\kappa,1)$ under $\mathbb Q$, so the last factor is $\kappa^2/2$. For the diffusion part, $dW=dW^*-\frac{\lambda_r}{\sigma_r}\sqrt r\,dt$ gives $\ln Z_{1,T}=-\frac{\lambda_r}{\sigma_r}\int_0^T\sqrt r\,dW^*+\frac{\lambda_r^2}{2\sigma_r^2}\int_0^Tr\,ds$, and the stochastic integral has zero $\mathbb Q$-mean.

\paragraph{Proposition \ref{prop:metric}.} Under $\mathbb Q$, $G_1$ and $G_2$ have mean zero, $\mathrm{Var}^{\mathbb Q}(G_1)=\mathrm{Var}^{\mathbb Q}(N_T)/(1+\lambda^\prime)^2=\lambda T/(1+\lambda^\prime)$, $\mathrm{Var}^{\mathbb Q}(G_2)=\mathbb E^{\mathbb Q}[N_T]=\lambda(1+\lambda^\prime)T$, and $\mathbb E^{\mathbb Q}[G_1G_2]=0$ because $\mathbb E^{\mathbb Q}[G_2\mid N_T]=0$. The form in the coordinates $\vartheta$ follows from the chain rule: with $\lambda^\prime=e^{\eta_f}-1$, the score with respect to $\eta_f$ is $\partial\log Z/\partial\eta_f=e^{\eta_f}G_1$, so that $g_{\eta_f\eta_f}=e^{2\eta_f}g_{11}=\lambda Te^{\eta_f}$, while $g_{\kappa\kappa}$ is unchanged. The gradient follows by differentiating $\mathcal K_C$.

\paragraph{Proposition \ref{prop:ident}.} With $s=\hat\lambda t$, $\mathbb Q(L_t<D)=e^{-s}\big(1+s\hat F(D)+\tfrac{s^2}{2}\hat F^{*2}(D)+O(s^3)\big)$, while $e^{-\ell t}=e^{-s}e^{s\hat F(D)}=e^{-s}\big(1+s\hat F(D)+\tfrac{s^2}{2}\hat F(D)^2+O(s^3)\big)$. Subtracting gives \eqref{eq:expansion}. Note that $\hat F^{*2}(D)\le\hat F(D)^2$. For a per-occurrence trigger, events exceeding $D$ form a thinned Poisson process with intensity $\ell(\theta)$, so the survival probability is exactly $e^{-\ell t}$.

\paragraph{Monte Carlo check of the closed forms.}\label{app:mc}
Table~\ref{tab:mc} compares the closed forms of Propositions~\ref{prop:KL} and~\ref{prop:metric} and the FFT survival probability with Monte Carlo estimates from $\XmcN\times10^6$ paths simulated under $\mathbb Q_{\theta^*}$, at the pooled optimum $\theta^*$ and horizon $T=17/4$. The metric entries are second moments of the scores $G_1,G_2$.

\begin{table}[htbp]
\centering\small
\caption{Closed forms versus Monte Carlo at the pooled optimum.}\label{tab:mc}
\begin{tabular}{lcccc}
\toprule
Quantity & closed form & Monte Carlo & s.e. & $z$\\
\midrule
$\mathcal K_C(\theta)$ & 0.32753 & 0.32845 & 0.00114 & +0.80\\
$g_{11}$ & 0.02395 & 0.02398 & 0.00006 & +0.58\\
$g_{22}$ & 0.30164 & 0.30265 & 0.00104 & +0.98\\
$g_{12}$ & 0.00000 & 0.00007 & 0.00018 & +0.41\\
$\mathbb Q(L_T<D)$ (FFT) & 0.82461 & 0.82434 & 0.00038 & -0.71\\
\bottomrule

\end{tabular}
\end{table}

\begin{figure}[htbp]
\centering
\includegraphics[width=0.95\linewidth]{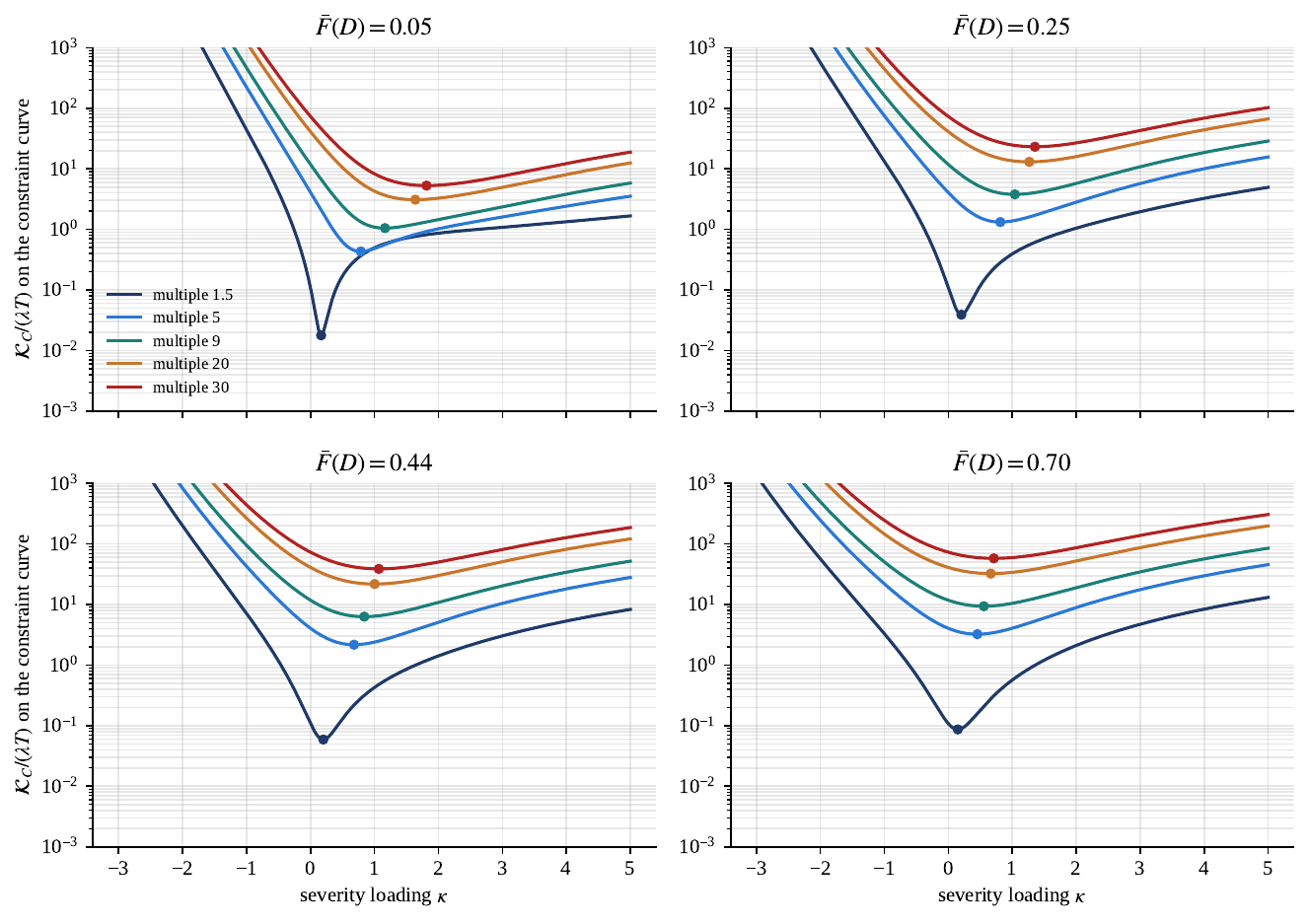}
\caption{Reduced objective $h(\kappa)=\mathcal K_C/(\lambda T)$ on the constraint curve $\{\ell(\theta)=\ell^*\}$ as a function of $\kappa$, for four physical exceedance probabilities $\bar F(D)$ and risk multiples $\ell^*/(\lambda\bar F(D))\in\{1.5,5,9,20,30\}$ (logarithmic scale); dots mark the minima.}
\label{fig:uniq}
\end{figure}

\paragraph{Condition \eqref{eq:foc}.} On the curve $\eta_f+\ln\Phi(\kappa-u_D)=\ln(\ell^*/\lambda)$, the Lagrange conditions for minimising $\mathcal K_C$ are $\lambda Te^{\eta_f}(\eta_f+\kappa^2/2)=\mu$ and $\lambda Te^{\eta_f}\kappa=\mu M(\kappa-u_D)$; eliminating $\mu$ gives \eqref{eq:foc}. Substituting the constraint, $\eta_f$ becomes a function of $\kappa$, and the problem reduces to minimising over $\kappa\in\mathbb R$ the one-dimensional objective
\[
h(\kappa)=\frac{\mathcal K_C}{\lambda T}=e^{\eta(\kappa)}\Big(\eta(\kappa)-1+\frac{\kappa^2}{2}\Big)+1,\qquad
\eta(\kappa)=\ln\frac{\ell^*}{\lambda}-\ln\Phi(\kappa-u_D),
\]
where $\ell^*/\lambda=m\,\bar F(D)$ for a risk multiple $m=\ell^*/(\lambda\bar F(D))$. The reduced objective $h$ tends to infinity as $\kappa\to\pm\infty$, so a minimiser exists. Uniqueness has been checked numerically but is not proved: for physical exceedance probabilities $\bar F(D)\in\{0.05,0.25,0.44,0.70\}$ and $120$ risk multiples in $[1.01,30]$ ($\XuniqN$ configurations), the reduced objective has exactly one local minimum on a fine grid of $\kappa\in[-10,10]$. Figure~\ref{fig:uniq} shows the reduced objective for representative configurations; in each case it has a single minimum.

\end{document}